\RequirePackage{fix-cm}
\documentclass[natbib,smallextended]{svjour3}   
\smartqed  
\usepackage{graphicx}
\usepackage{amssymb}
\usepackage{amsmath}
\usepackage{appendix}
\usepackage{upgreek}
\usepackage[normalem]{ulem}
\usepackage{rotating}
\usepackage{footnote}
\usepackage[caption=false]{subfig}
\usepackage{wrapfig}
\usepackage[dvipsnames]{xcolor}
\usepackage{comment}
\usepackage[colorlinks=true,citecolor=blue,urlcolor=blue,breaklinks]{hyperref}
\newcommand{\aofrb}{FRB~20121102A}

\newcommand{\edit}[1]{\textcolor{black}{#1}}

\def\cite{\citep}

\journalname{The Astronomy and Astrophysics Review}
\begin{document}

\title{Fast radio bursts at the dawn of the 2020s}


\author{E. Petroff         \and
        J. W. T. Hessels \and
        D. R. Lorimer 
}


\institute{E. Petroff \at
  Anton Pannekoek Institute, University of Amsterdam, Science Park 904, 1098 XH, Amsterdam, The Netherlands\\
  McGill Space Institute, McGill University, 3550 rue University, Montr\'eal, QC H3A 2A7, Canada\\
              \email{e.b.petroff@uva.nl}           
\and
J.~W.~T.~Hessels \at
    Anton Pannekoek Institute, University of Amsterdam, Science Park 904, 1098 XH, Amsterdam, The Netherlands\\
    ASTRON, Netherlands Institute for Radio Astronomy, Oude Hoogeveensedijk 4, 7991 PD, Dwingeloo, The Netherlands
\and        
D.~R.~Lorimer \at
    Department of Physics and Astronomy, West Virginia University, PO Box 6315, Morgantown, WV, USA\\
    Center for Gravitational Waves and Cosmology, West Virginia University, Chestnut Ridge Research Building, Morgantown, WV, USA
}

\date{Received: date / Accepted: date}

\maketitle

\begin{abstract}
Since the discovery of the first fast radio burst (FRB) in 2007, and their confirmation as an abundant extragalactic population in 2013, the study of these sources has expanded at an incredible rate. In our 2019 review on the subject we presented a growing, but still mysterious, population of FRBs -- 60 unique sources, 2 repeating FRBs, and only 1 identified host galaxy. However, in only a few short years new observations and discoveries have given us a wealth of information about these sources. The total FRB population now stands at over 600 published sources, 24 repeaters, and \edit{19} host galaxies. Higher time resolution data, sustained monitoring, and \edit{precision localisations} have given us insight into repeaters, host galaxies, burst morphology, source activity, progenitor models, and the use of FRBs as cosmological probes. The recent detection of a bright FRB-like burst from \edit{the} Galactic magnetar SGR~1935+2154 provides an important link between FRBs and magnetars.  There also continue to be surprising discoveries, like periodic \edit{modulation of} activity from repeaters and the localisation of one FRB source to a relatively nearby globular cluster associated with the M81 galaxy. In this review, we summarise the exciting observational results from the past few years.  \edit{We also highlight} their impact on our understanding of the FRB population and proposed progenitor models. We build on the introduction to FRBs in our earlier review, update our readers on recent results, and discuss interesting avenues for exploration as the field enters a new regime where hundreds to thousands of new FRBs will be discovered and reported each year. 

\end{abstract}

\setcounter{tocdepth}{3} 
\tableofcontents

\section{Introduction}\label{sec:intro}

Research into understanding the origins, mechanisms and applications
of fast radio bursts (FRBs) is currently one of the most rapidly evolving areas in astrophysics. As a phenomenon, FRBs are characterised as short-duration (\edit{sub-second}), relatively broad-band pulses which exhibit the frequency-dependent arrival time delay due to dispersion that is well known for Galactic radio pulsars \citep[see, e.g.][]{lorimer_2012_hpa}. Unlike pulsar emission, \edit{however}, FRB pulses show much larger dispersive delays, consistent with an extragalactic origin; they also sometimes show characteristic time-frequency drifts, even after dedispersion \citep{hessels_2019_apjl}. In our \edit{previous} review \citep[][hereafter PHL19]{petroff_2019_aarv} we \edit{provided} an extensive introduction to FRBs \edit{starting with} their discovery \citep{lorimer_2007_sci}, and \edit{detailed} their 
emergence as a cosmological source population whose origins are still poorly understood. Further insights, and a much more detailed examination of propagation effects, can be found in \citet{cordes_2019_araa}.

\begin{figure}[htb]
    \centering
    \includegraphics[width=\linewidth]{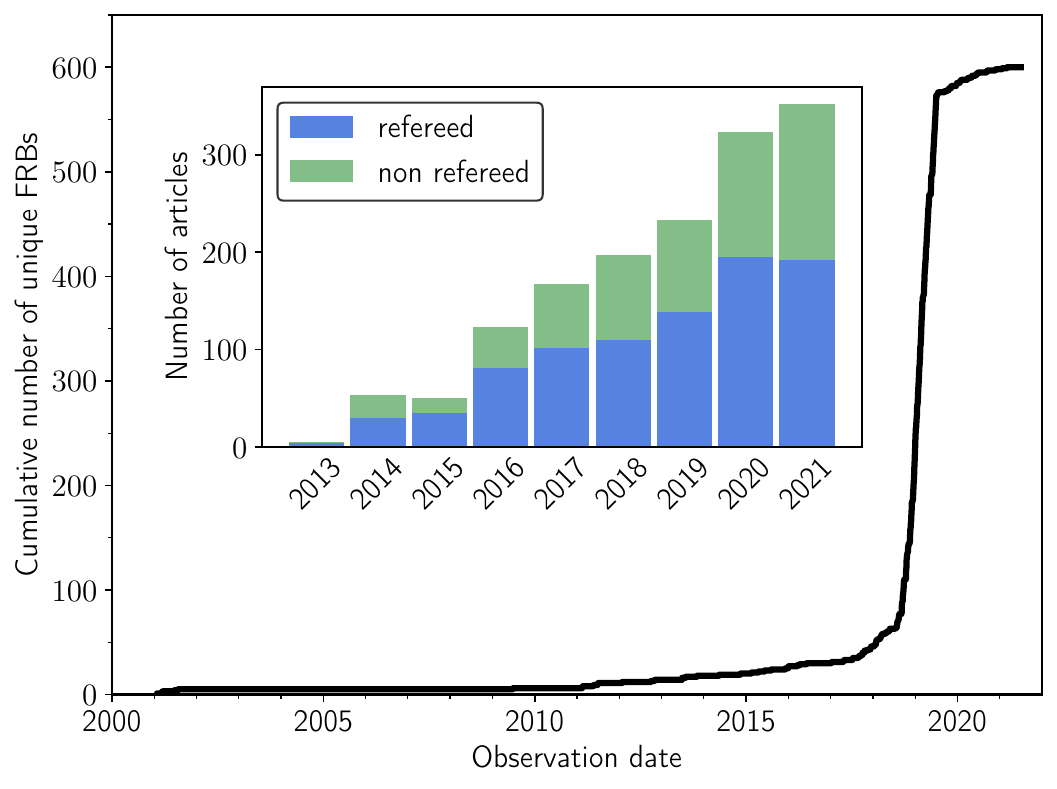}
    \caption{Cumulative number \edit{of published, unique FRB sources} as a function of their arrival time at the Earth.  \edit{Some of these predate the Lorimer burst \citep[FRB~010724;][]{lorimer_2007_sci} because they were subsequently found in other archival data from the Parkes radio telescope.} The sample is dominated by the recent publication of the first CHIME/FRB catalogue \edit{of FRBs} \citep{chime_2021_arxiv_arxiv210604352}.  \edit{The leveling off in recent years reflects the rate at which new FRB sources are being published; they continue to be discovered at an ever increasing rate.} Inset: publications with `fast radio burst' in the abstract since 2013, \edit{the year in which the term was coined in \citet{thornton_2013_sci}}.  Source: NASA/ADS, \edit{TNS and FRBSTATS}.  \edit{Figure courtesy Mark Snelders.}}
    \label{fig:numbers}
\end{figure}

Here, in this new, \edit{updated} review, we provide a survey of the main developments from the last 2--3 years. Significant breakthroughs have been made in this period thanks to the emergence of new observational facilities, as well as a growing and dynamic group of active researchers with a wide range of backgrounds. Fig.~\ref{fig:numbers} illustrates the large effect that this has had on the discovery rate of FRBs and research presented in the literature\footnote{See also FRBSTATS, \url{https://www.herta-experiment.org/frbstats}}.

\begin{figure}[htb]
    \centering
    \includegraphics[trim={0 3cm 0 3cm},clip,width=\linewidth]{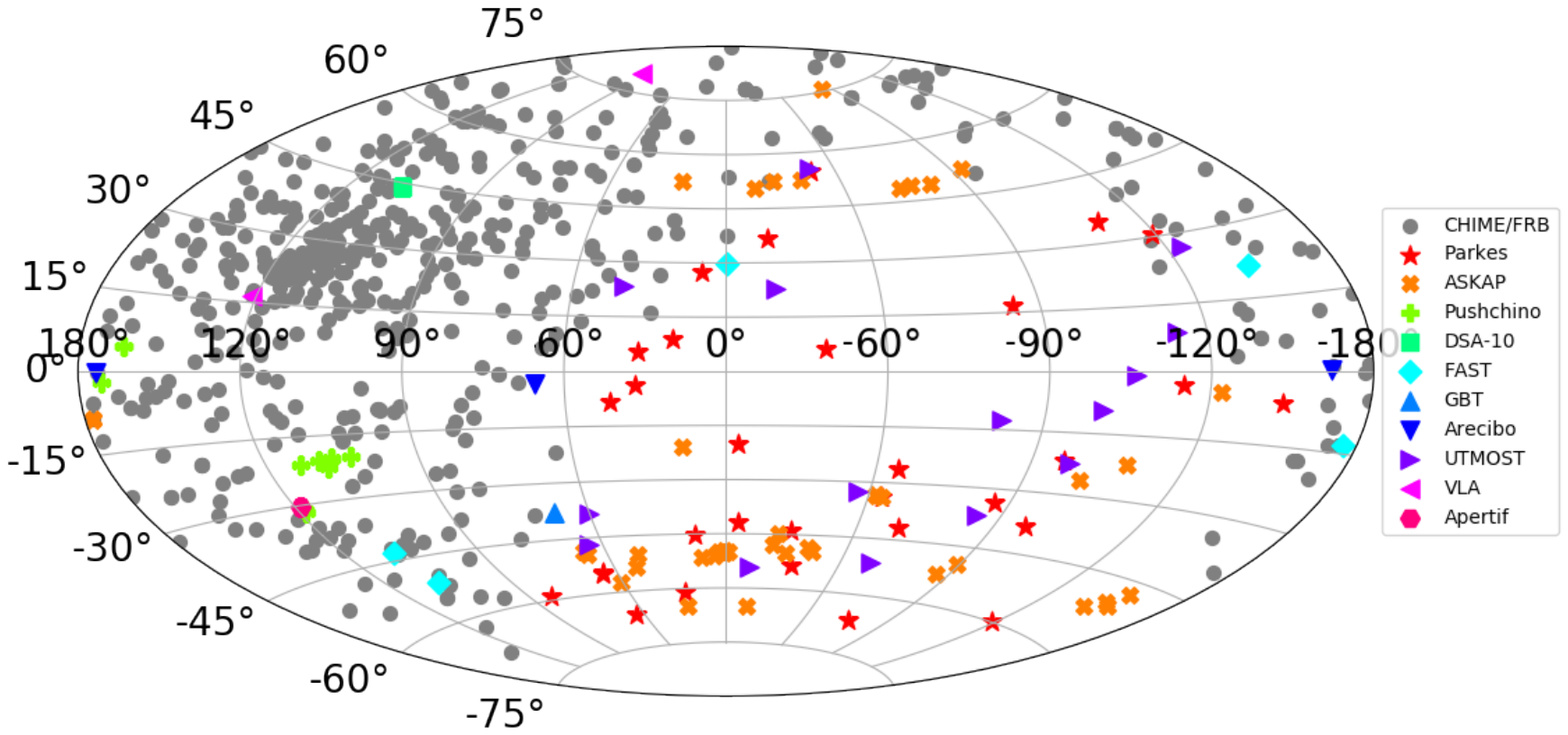}
    \caption{Aitoff projection showing the \edit{currently known} sample of FRBs in Galactic coordinates. The FRBs shown here are predominantly those found by CHIME/FRB, which is known to have significant variations in sensitivity over the sky and only views the northern celestial hemisphere. \edit{When corrected for non-uniform sky coverage, it has been shown that the underlying distribution is isotropic, as expected for a cosmological population.}}
    \label{fig:aitoff}
\end{figure}

\begin{figure}
\centering
\subfloat[DM excess histogram\label{fig:DMHist}]{\includegraphics[width=0.7\textwidth]{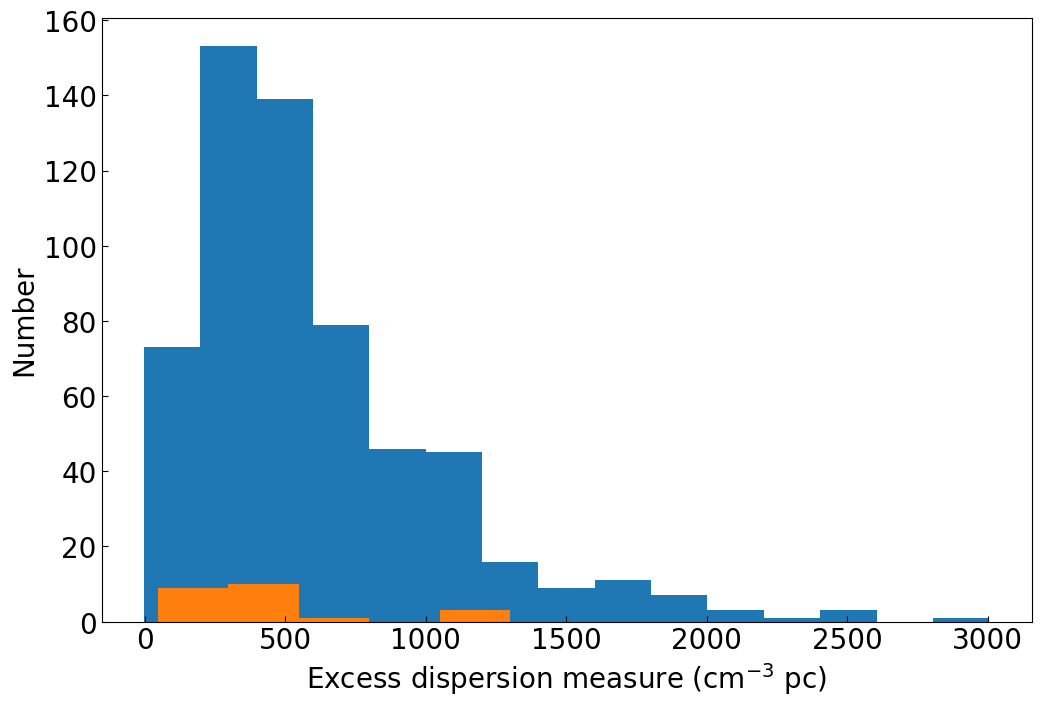}}
\hfill
\subfloat[Pulse duration histogram\label{fig:widthHist}]{\includegraphics[width=0.7\textwidth]{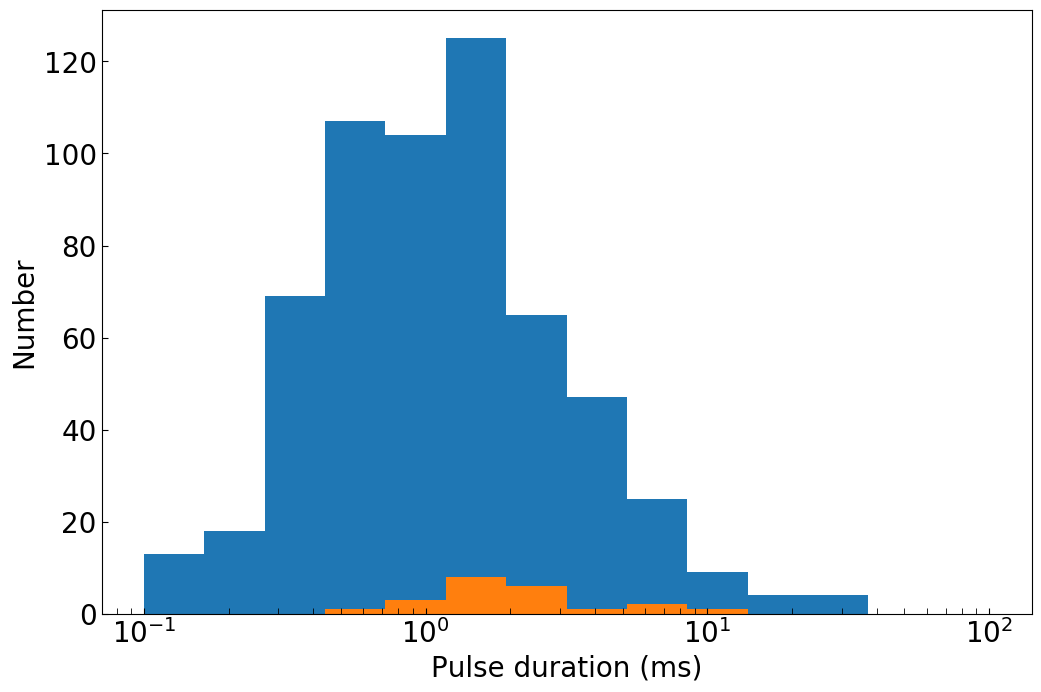}}
\vfill
\subfloat[Pulse duration versus DM\label{fig:DMWidth}]{\includegraphics[width=1.0\textwidth]{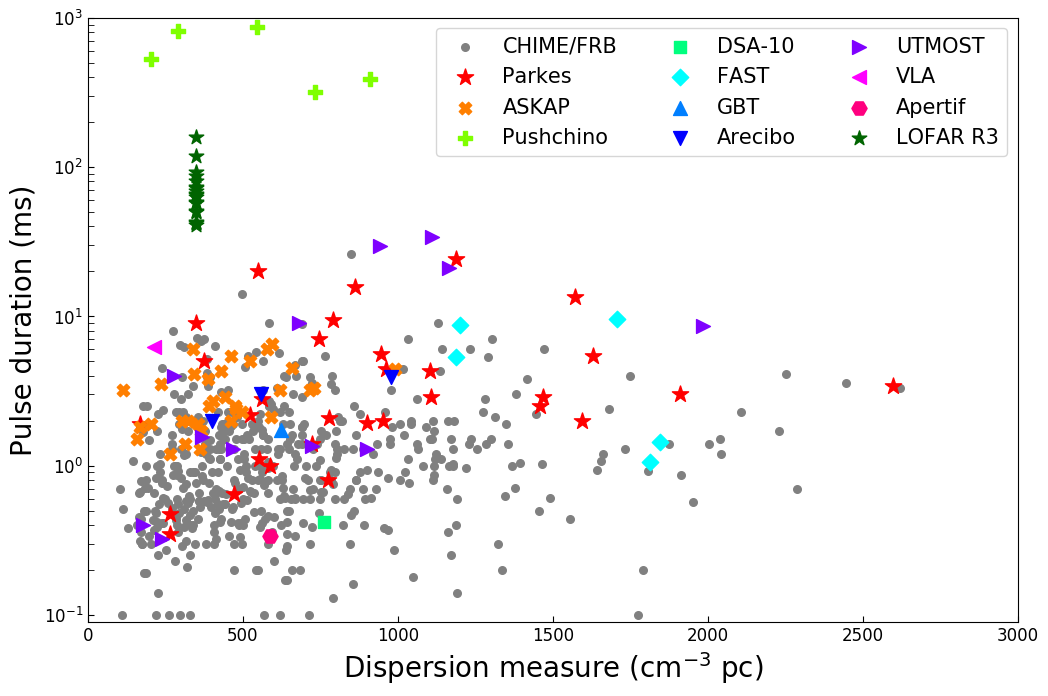}}
\caption{
Selected observed properties of the \edit{currently known} sample of FRBs for one-off bursts (blue) and the first-detected bursts from observed repeaters (orange). (a) The distribution \edit{of reported DMs}. (b) The distribution \edit{of reported widths}. (c) The \edit{reported} widths versus DM for all one-offs and first repeats from different detection instruments; here the repeat bursts from FRB~20180916B (R3) detected with LOFAR from \citet{pleunis_2021_apjl} are also plotted because they illustrate the large widths seen at low radio frequencies (\edit{likely} dominated by scattering \edit{and time-frequency drift} in this case). 
\label{fig:Distributions}}
\end{figure}

Below we briefly list the key areas to be discussed in this review:
\begin{itemize}
    \item {\bf \edit{Overall} sample size:} In PHL19, the published sample of FRBs totalled around 60. Thanks to the torrent of discoveries since that time, most notably with the release of the first catalogue from the Canadian Hydrogen Intensity Mapping Experiment Fast Radio Burst project (CHIME/FRB), the currently published sample now exceeds 600 unique sources. \edit{The observed properties of} this population, which \edit{supersede} those presented in PHL19, are shown in Fig.~\ref{fig:aitoff} \edit{and} Fig.~\ref{fig:Distributions}.  As is often the case in burgeoning research fields, such a quantum leap in the sample size inevitably leads to fundamental breakthroughs in our understanding. 
    \item {\bf Repeating FRBs:} Since PHL19, where only two repeating FRBs were known, the sample of published repeating sources has now grown to 24. CHIME/FRB in particular --- with its \edit{wide field-of-view,} repetitive coverage of the sky and good sensitivity --- has been very effective at finding these sources. \edit{Considering these, along with the apparently one-off FRBs,} we discuss the evidence for and against there being multiple source populations \edit{with distinct physical origins}. 
    \item {\bf Periodic activity in repeating FRBs:} Due to their \edit{sporadic and generally unpredictable activity}, monitoring repeating FRBs is a \edit{time intensive} exercise. Nevertheless, thanks to the accumulation of long-term observing campaigns at multiple observatories, we now know that there is an underlying periodicity in activity windows for at least two repeating FRBs. Additionally, sub-second periodicity between sub-bursts within a single event has now been seen for at least one FRB.  We place these results within the context of source models for FRBs. 
    \item {\bf \edit{Burst} energy and luminosity distributions:} The \edit{first-known} repeater, \aofrb\footnote{\edit{This source was originally named FRB~121102.}  The FRB community now follows the Transient Name Server (TNS) naming convention of `FRB YYYYMMDDX'\edit{, which includes} the year, month, day, and \edit{and an additional letter to distinguish events that were reported on the same day}. See \url{https://www.wis-tns.org}}, has now been monitored extensively with a number of high-sensitivity instruments and has by far the largest number of \edit{bursts} recorded. Recently, several groups have announced a significant increase in the number of \edit{bursts detected} from \aofrb. This sample, coupled with an accurate measurement of the source redshift now provides a relatively complete view of the \edit{burst} energy \edit{and (isotropic equivalent) luminosity} distribution \edit{of} this \edit{source}. We also discuss progress in our understanding of the FRB luminosity function for the population as a whole. \edit{The recent discoveries of repeating FRB sources associated with relatively nearby galaxies present} an opportunity to study the faint end of the FRB luminosity distribution.
    \item {\bf Host galaxy identifications:} Since PHL19, at which point only one host galaxy identification \edit{(for the first-known repeater)} was conclusively made, the sample of FRBs with accurate positions that allow for robust host galaxy association and redshift determination\footnote{For an up-to-date list of FRBs with host galaxy identifications, see \url{https://frbhosts.org}.}  has increased to \edit{19}. Among the new localizations are some exceptionally nearby FRBs, like a repeater associated with a globular cluster in the M81 galactic system, which provide strong constraints on multi-wavelength emission. The overall increase in the number of \edit{host galaxy} identifications, \edit{in particular for one-off FRBs}, is largely due to the interferometric capabilities and sky coverage of ASKAP \edit{(about a dozen one-offs localised)}. In addition, very long baseline interferometry (VLBI) with the European VLBI Network (EVN) is also providing milli-arcsecond localisations for an increasing number of \edit{repeating} sources \edit{(six such localisations published or in preparation)}, \edit{and gives} enough precision to match high-resolution {\it Hubble Space Telescope} images of the hosts. These studies indicate that FRBs are produced in a wide variety of host galaxy types; \edit{a larger sample is needed to investigate potential systematic differences between the hosts and local environments of repeaters and apparent one-off FRBs}. 
    \item {\bf Probing the intergalactic medium:} \edit{Since PHL19, the emerging sample of FRBs with known redshift has been} used to establish the so-called `Macquart relationship'\footnote{\edit{Named in memory of our colleague and friend, J.-P. Macquart.}} between dispersion measure (DM) and redshift.  \edit{This} has led to a best-ever quantification of the intergalactic baryon content.
    \item {\bf FRB-like pulses from a Galactic magnetar:} The diversity of \edit{FRB host galaxy types}, mentioned above, includes spiral galaxies that are analogues of the Milky Way (e.g., the host of the nearby repeater FRB~20180916B).  \edit{Shortly before the discovery of the Lorimer burst, it was also found} that Galactic magnetars \edit{can} have radio outbursts in which bright pulses \edit{are} detected. In 2020, the Galactic magnetar SGR\footnote{\edit{SGR stands for soft gamma repeater.}}~1935+2154 went into a highly 
    active state in which a mega-Jansky \edit{burst} was detected by STARE2 and CHIME/FRB. When placed into context with extragalactic FRBs, these detections are consistent with being part of the low end of a broad luminosity function, and provide strong circumstantial evidence linking magnetars \edit{and} FRBs.
\end{itemize}

These rapid advances within the FRB research community \edit{demonstrate the power of} modern time-domain astrophysics, which has also seen significant developments in gravitational wave detection with ground-based instruments \citep[for a recent review, see][]{miller_2019_natur} and large-scale surveys of the sky at other wavebands
\citep[see, e.g.,][]{bloemen_2016_spie,bellm_2019_pasp,nir_2021_pasp}. Together, these efforts are establishing the foundations for a 
cosmic census of compact object source populations. Further discussions of recent
FRB results can be found in \citet{zhang_2020_natur},
\citet{chatterjee_2021_ag}, \citet{bhandari_2021_univ} and \citet{xiao_2021_scpma}.  \edit{Summaries of the recent FRB annual conferences, FRB~2020 \citep{keane_2020_natas} and FRB~2021 \citep{hessels_2021_natas}, are also available}.

\section{Recent observational breakthroughs}\label{sec:observations}

Both the discovery of more FRBs and deeper characterisations of their properties are key to understanding their physical origins \edit{-- as well as} for using them as astrophysical \edit{and cosmological} probes.  The past few years, since our original FRB review appeared, have provided important new observational insights \edit{at an accelerating pace, along with} some major surprises.  \edit{We now discuss these in more} detail.

\subsection{Having lots of them}\label{sec:2_1}

The global effort to study FRBs has led to an explosion in the number of experiments running worldwide and, as a natural consequence, the number of known FRBs.  The known population in the literature\footnote{See the Transient Name Server (\url{https://www.wis-tns.org}), and a summary of the \edit{statistics} at FRBSTATS (\url{https://www.herta-experiment.org/frbstats}).} now stands at $\sim600$ one-off events and two dozen repeaters, thanks in large part to CHIME/FRB \citep{chime_2018_apj,chime_2021_arxiv_arxiv210604352}, whose unprecedented $\sim 200$\,sq.~deg. field of view and nearly continuous operation allow for the discovery of several new FRBs per day\footnote{\edit{These are now also being announced in near real time using VOEvents, enabling rapid multi-wavelength and multi-messenger follow-up.}}.  Repeating FRBs currently account for $\sim4$\% of the known population, but many repeaters have only been seen twice.  This suggests that many currently one-off sources may be seen to produce a second (or third, etc.) burst, given enough follow-up.

Nonetheless, the large sample of known FRBs is starting to show trends that suggest sub-populations based on burst morphology and spectra.  A paradigm for repeating FRBs is appearing to emerge: given enough S/N and time resolution, \edit{they} often show a characteristic downward drift of sub-bursts in frequency
\citep[][this is colloquially referred to as the `sad trombone' effect]{hessels_2019_apjl,fonseca_2020_apjl}; $\sim100$\% linear polarisation and negligible circular polarisation \citep{nimmo_2021_natas}, but not always \citep{hilmarsson_2021_mnras,kumar_2021_arxiv}; and two sources show a periodicity in their activity level \citep{chime_2020_natur_582,rajwade_2020_mnras,cruces_2021_mnras}. \citet{pleunis_2021_apj} study the sample of bursts from the first CHIME/FRB catalogue and find that repeaters are typically wider in time but narrower in bandwidth compared to one-offs. It thus appears that there are likely at least two types of FRBs, and that one-offs are not just simply less active sources.  Whether this demonstrates different types of sources, or whether it can be accommodated by a different (and more sporadic) emission mechanism from the same type of source, is unknown.  

\begin{figure}
    \centering
    \includegraphics[width=0.9\columnwidth]{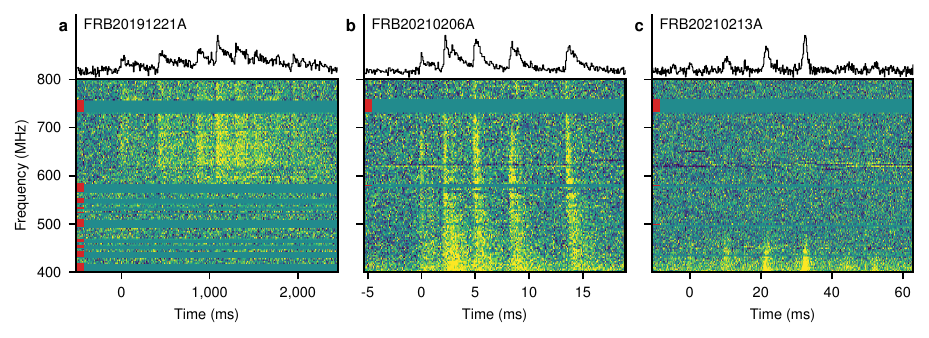}
    \caption{\edit{Periodic and quasi-periodic bursts detected by CHIME/FRB.  FRB~20191221A (left)
    is a remarkable event, lasting for nearly 3 seconds and showing 9 peaks separated by a strict period of 216.8\,ms.  It also shows a high level of scattering.  The other two events, FRB~20210206A and FRB~20210213A, show quasi-periodicities at 2.8\,ms and 10.7\,ms, respectively --- though the statistical significance of these periods are marginal.  We note that the sub-bursts of repeaters are sometimes also quasi-periodic \citep{hessels_2019_apjl}.  Figure from \citet{chime_2021_arxiv_arxiv210708463}.}}
    \label{fig:periodic_bursts}
\end{figure}

As the known FRB population continues to increase rapidly, further sub-populations may become apparent.  \edit{FRB~20191221A, the only known FRB to show a strict periodicity between sub-bursts (which are separated by 216.8\,ms, see Fig.~\ref{fig:periodic_bursts}), appears to be a rare type of event and one can speculate that it is different in origin compared to the vast majority of the observed FRB population.}

Large samples of bursts have also become available for some individual repeaters, especially \aofrb, FRB~20180916B and FRB~20201124A.  These allow us to study their individual burst energy distributions and average spectra, as well as how their activity and properties change with time (as discussed below).  The isotropic equivalent energy distribution of \aofrb\ is multi-modal, and an apparent peak in the distribution is found at lower energies \citep{li_2021_natur}.   
This shows that repeaters can produce multiple types of bursts, some of which arguably appear more similar to those seen from one-off FRBs.  The large sample of bursts found by FAST \citep{li_2021_natur} and Arecibo \citep{hewitt_2021_arxiv} has yet to demonstrate any strict or quasi-periodicity in the burst arrival times, like what is seen from pulsars.  \edit{Nonetheless, the wait time between sub-bursts is characteristically about a few ms \citep{hessels_2019_apjl,li_2021_natur}, and there is also a broad peak in the wait times between closely spaced bursts at roughly a few tens of ms \citep{hewitt_2021_arxiv} --- in addition to a dominant peak related to the general burst rate.}  The large number of bursts \edit{per unit time} also means that the available energy reservoir of the central engine is not easily exhausted.  \edit{Indeed, \aofrb\ has been observed as an active repeater for now close to a decade \citep{spitler_2014_apj,spitler_2016_natur}, and this constrains the lifetime of such sources.}

Overall, the much larger sample of known sources, and burst distributions for individual sources, is allowing for increasingly meaningful population studies \citep{james_2022_mnras_1,james_2022_mnras_2}.  These are beginning to better quantify how well the observed FRB population traces star formation as a function of redshift and allows for comparison of the volumetric rate with those of other types of astrophysical events (e.g., supernovae, binary mergers of compact objects) and sources \citep[see \S~\ref{sec:3_5}; also][]{ravi_2019_natas,zhang_2021_mnras}.

\edit{The} applications of FRBs to broader problems in astrophysics and cosmology are greatly aided by these large samples (see \S~\ref{sec:probes}).  At the same time, we have entered the regime in which there is insufficient person-power and telescope resources to follow-up every FRB exhaustively.

\subsection{Knowing where they come from}\label{sec:2_2}

\edit{Close to two dozen} FRBs have now been localised to (sub-)arcsecond precision and robustly associated with a host galaxy.  Interestingly, these studies have shown that FRBs can be hosted in a wide range of galaxy types and environments, though some trends are possibly emerging \citep[see \S~\ref{sec:3_2}; also][]{heintz_2020_apj,mannings_2021_apj,li_2020_apjl,bhandari_2021_arxiv}.  The precisely localised repeaters have been \edit{shown, in some cases, to be close to} star-forming regions \edit{in their host galaxies} \citep{bassa_2017_apjl,marcote_2020_natur,fong_2021_apjl,ravi_2021_arxiv,piro_2021_aa,nimmo_2021_arxiv_arxiv211101600}.  However, in the cases of \aofrb\ \citep{bassa_2017_apjl} and FRB~20180916B, \citep{tendulkar_2021_apjl} \edit{there is an offset of $\sim$200--250\,pc between} the FRB source and the closest peak of local star formation \edit{when comparing the milliarcsecond localisation from the EVN with {\it Hubble Space Telescope imaging}}.  \edit{Additional localisations are needed to determine whether this is a common trend; does it indicate that repeaters are older than one might naively expect, or do they come from runaway OB stars that collapsed to a neutron star relatively far from their birthplace?}  Conversely, some non-repeaters have been found in galactic hosts with very low star-formation rates, and even in the outskirts of such galaxies \citep{heintz_2020_apj,mannings_2021_apj,bhandari_2021_arxiv}.  This has led to the hypothesis that FRBs may be \edit{magnetars} formed via a variety of channels including core collapse supernovae, binary mergers and accretion-induced collapse \citep{margalit_2019_apj,gourdji_2020_mnras}.  \edit{They may also be older systems than assumed in young magnetar models --- e.g., high-mass binaries featuring a neutron star \citep{tendulkar_2021_apjl}}.

\begin{figure}
    \begin{tabular}{cc}
    \includegraphics[width=0.46\columnwidth]{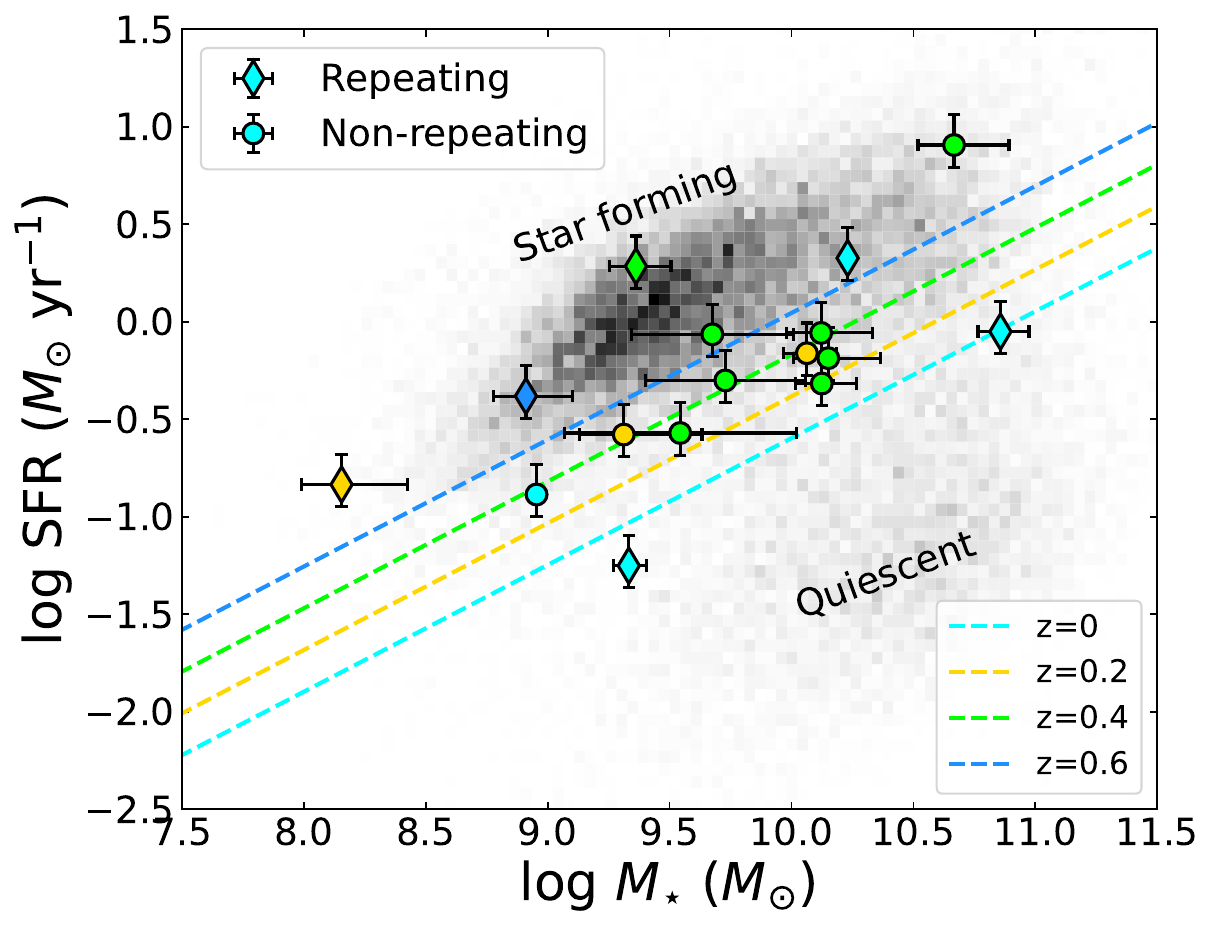}
    \includegraphics[width=0.45\columnwidth]{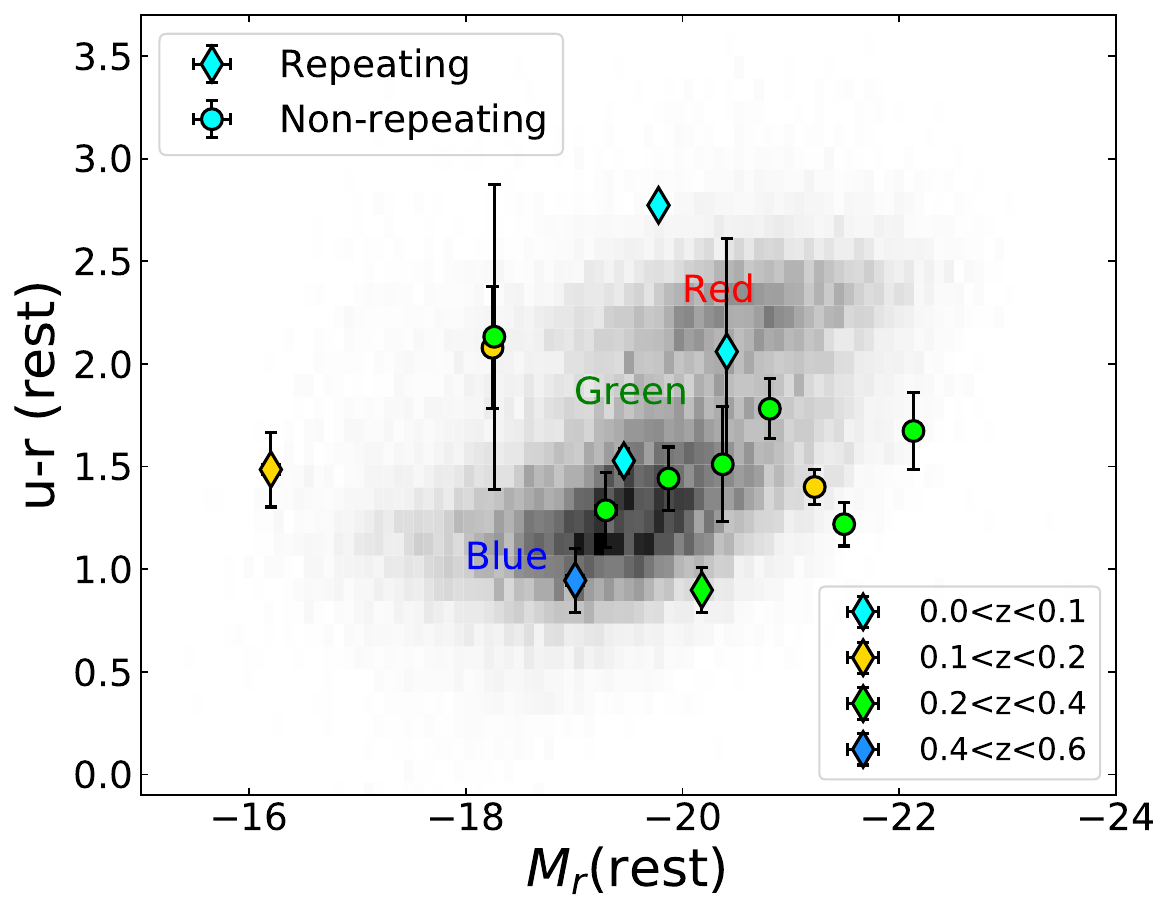}
    \end{tabular}
    \caption{\edit{{\it Left}: Star-formation rate versus stellar mass for the host galaxies of 16 FRBs with robust host associations.  These are compared with a population of galaxies at $z < 0.6$.  {\it Right}: Rest-frame color-magnitude diagram of the host galaxies of 15 FRBs (all those in the left panel, with the exception of FRB~20171020A).  These are compared with the same galaxy sample as in the left panel.  Figure from \citet{bhandari_2021_arxiv}.}}
    \label{fig:bhandari_hosts}
\end{figure}

\begin{figure}
    \centering
    \includegraphics[width=0.9\columnwidth]{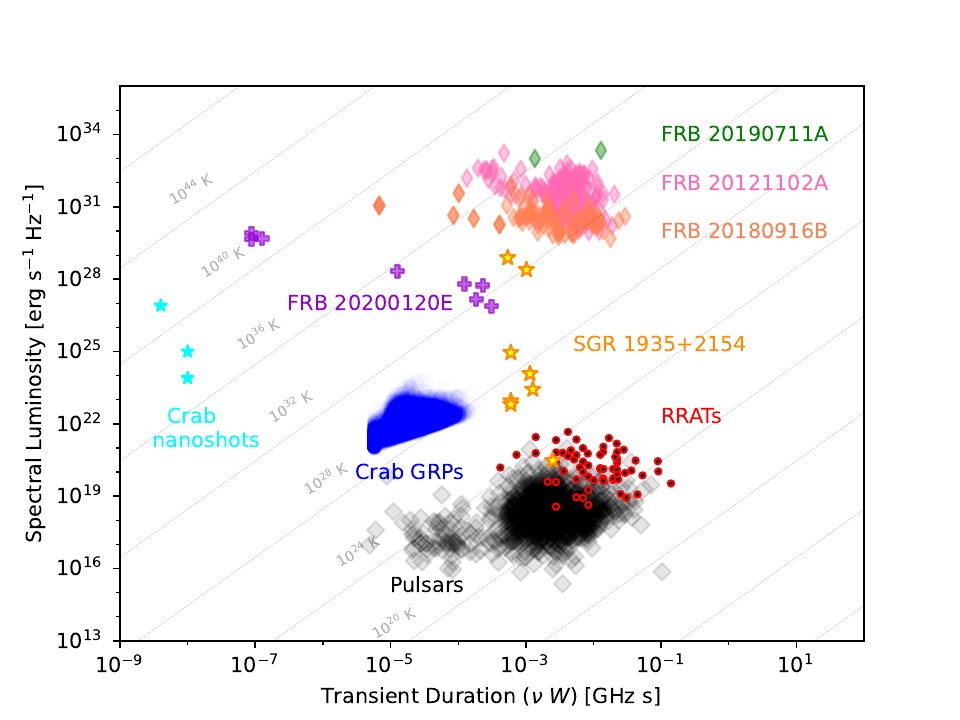}
    \caption{Radio transient phase space, focusing on the sub-second coherent radio flashes seen from Galactic pulsars \edit{(including the giant radio pulses, GRPs, and `nanoshots' seen from the Crab), rotating radio transients (RRATs)} and magnetars, as well as the extragalactic fast radio bursts. The pulsar luminosities plotted here are averaged and pulse-to-pulse variations can be significant.  \edit{For both the pulsars and FRBs, the plotted luminosities are isotropic equivalent, because the beaming angle is unknown.}  Courtesy K.~Nimmo, \edit{adapted from} \citet{nimmo_2021_arxiv_arxiv210511446}.}
    \label{fig:tps_zoom}
\end{figure}

While the majority of localised FRBs are \edit{located in galaxies at Gpc distances}, we are also fortunate to have detected a putative FRB in our own Milky Way galaxy.  On 28 April 2020, CHIME/FRB and STARE2 simultaneously detected\footnote{\edit{To within the dispersive delay between their observing bands.  Also, note that CHIME/FRB detected two sub-bursts in their 400--800\,MHz band, whereas only one of these components was visible at 1.4\,GHz with STARE2.}} an exceptionally bright (1.5\,MJy~ms at 1.4\,GHz) burst from the Galactic magnetar SGR~1935+2154 \citep{chime_2020_natur_587,bochenek_2020_natur}.  This burst, referred to by some in the literature as FRB~200428, has a spectral luminosity that is only a factor of 30 less compared to the weakest known bursts from FRB~20180916B, which until recently was the nearest-known extragalactic FRB source, at $\sim 150$\,Mpc \citep{marcote_2020_natur}.  The MJy-ms event from SGR~1935+2154 is actually higher in spectral luminosity compared to some bursts seen from an exceptionally nearby FRB found to be associated with M81 (see below).  SGR~1935+2154 has shown several other bursts, one with a fluence of only 60\,mJy~ms \citep[detected by FAST; ][]{zhang_2020_atel}, and others with fluences of about 10--100\,Jy~ms \citep{kirsten_2021_natas}.  Together, these bursts span $\sim 7$ orders-of-magnitude in spectral luminosity, and the burst rate is remarkably similar across this range \citep{kirsten_2021_natas}. This bridges the luminosities of Galactic radio pulse emitters and the extragalactic FRBs (Fig.~\ref{fig:tps_zoom}).
\edit{As recently shown by \citet{lin_2020_natur}, however, with FAST observations of SGR~1935+2154 taken during a subsequent period of X-ray activity, no accompanying FRB-like radio pulses were detected in spite of a noise floor that is 8 orders of magnitude lower. This null result suggests that associations between SGR bursts and FRBs are very rare.}

The exceptionally active FRB~20201124A has been localised by several interferometers, providing the best test yet of localisation accuracy, both within and between experiments \citep{ravi_2021_arxiv,fong_2021_apjl,nimmo_2021_arxiv_arxiv211101600}.  Interestingly, \edit{the FRB source} was shown to be coincident with a persistent radio source.  However, VLBI observations with EVN and VLBA have demonstrated that this source is not compact \citep{ravi_2021_arxiv,nimmo_2021_arxiv_arxiv211101600}, unlike the persistent radio counterpart to \aofrb.  
\edit{A VLA detection at 22\,GHz demonstrates that this radio source is very likely due to star formation \citep{piro_2021_aa}; future {\it Hubble Space Telescope} observations will better quantify where FRB~20201124A is located with respect to local knots of star-formation, by exploiting its milliarcsecond localisation \citep{nimmo_2021_arxiv_arxiv211101600}.}
Bursts from FRB~20201124A can be exceptionally bright ($>100$\,Jy~ms), and coupled with its high activity rate, this has allowed for \edit{detections} with $\sim25$-m class \citep{kirsten_2021_atel} and \edit{even} smaller radio dishes \citep{farah_2021_atel}, \edit{as well as one dish that is operated by amateur radio astronomers \citep{herrmann_2021_atel}}.

The CHIME/FRB discovery of a low-DM repeater, \edit{FRB~20200120E}, in the direction of \edit{the} nearby M81 galaxy \citep{bhardwaj_2021_apjl_910} has provided the opportunity to study \edit{an} extragalactic FRB from nearby.  \citet{kirsten_2021_arxiv} localised this source to a globular cluster in the M81 system at 3.6\,Mpc, thereby ruling out that it is a Galactic halo source.  The surprising association with an old globular cluster rules out a young source formed through core collapse of a massive star, and \edit{possibly} points towards a young, \edit{highly magnetised neutron star} formed via accretion-induced collapse or binary merger \citep{kirsten_2021_arxiv,lu_2022_mnras,kremer_2021_apjl}.  Alternatively, the bursts could be coming from an extreme millisecond pulsar \citep[][though \citealt{lu_2022_mnras} argue against this]{kremer_2021_apjl}, or \edit{an} accreting neutron star or black hole \citep{sridhar_2021_apj}.  \edit{The low burst luminosities and short burst durations of FRB~20200120E \citep{nimmo_2021_arxiv_arxiv210511446} suggest that, perhaps, it is not representative of the broader population of repeating FRB sources.}

\begin{figure}
    \centering
    \includegraphics[width=0.9\columnwidth]{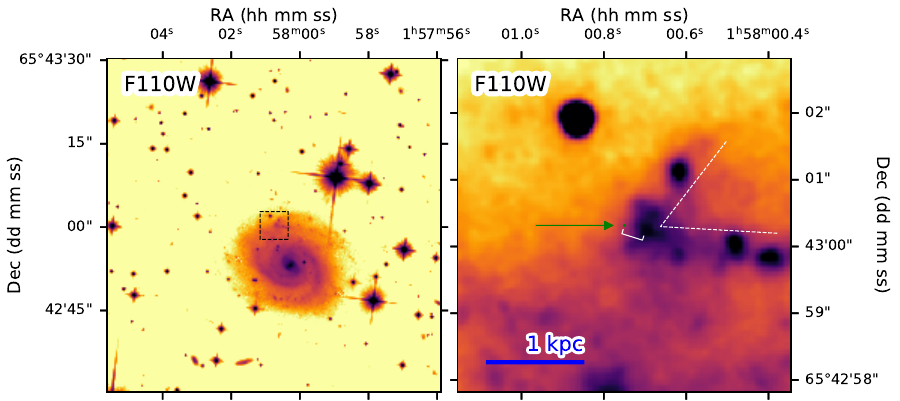}
    \caption{\edit{{\it Hubble Space Telescope} images of the host galaxy of FRB~20180916B.  The left panel shows the full galaxy and its surroundings, while the right panel provides a zoom-in (defined by the dashed square in the left panel).  The FRB position, as determined from EVN, is offset by $\sim250$\,pc from the closest peak of local star formation.  Figure adapted from \citet{tendulkar_2021_apjl}.}}
    \label{fig:r3_host}
\end{figure}

\edit{Even more recently, \citet{bhardwaj_2021_apjl_919} presented the likely host of FRB~20181030A.  Using CHIME/FRB \edit{voltage} data, they were able to localise the source to a sky region of a few square arcminutes.  Despite the relatively coarse localisation precision, the exceptionally low DM of the source and overlap with the prominent galaxy NGC~3252 suggest that the association is robust.  NGC~3252 is a star-forming spiral galaxy at a distance of $\sim20$\,Mpc, placing FRB~20181030A at an intermediate distance between the other well-localised repeaters FRB~20200120E ($\sim3.6$\,Mpc) and FRB~20180916B ($\sim150$\,Mpc).  Importantly, \citet{bhardwaj_2021_apjl_919} argue that young millisecond magnetars alone are insufficient to explain the observed volumetric rate of repeating FRBs.}

\edit{Most recently, the FAST-discovered repeater FRB~20190520B has been shown to be coincident with a compact persistent radio source in a dwarf host galaxy \citep{niu_2021_arxiv}.  This makes FRB~20190520B a close analogue to the original repeater (\aofrb) though, interestingly, the local and host DM contribution of $\sim900$\,cm$^{-3}$~pc is much higher than any other localised FRB (see also \S~\ref{sec:4_1} and Fig.~\ref{fig:DMz}).  This may indicate that FRB~20190520B is a particularly young source, or at least that it was formed in an exceptionally dense environment.  Active repeaters have been found both {\it with} \citep{chatterjee_2017_natur,marcote_2017_apjl,niu_2021_arxiv} and {\it without} \citep{marcote_2020_natur,kirsten_2021_arxiv,nimmo_2021_arxiv_arxiv211101600} a compact persistent radio source (which is unrelated to star formation).  \citet{law_2021_arxiv} discuss how such sources may comprise as much as 1\% of the compact, luminous radio sources detected in the local Universe.  Their nature remains unclear, however: e.g., do they represent magnetar wind nebulae, or perhaps a nearby accreting massive black hole \citep{michilli_2018_natur,eftekhari_2019_apjl}?  FRBs from these compact persistent radio sources may be discoverable in targeted searches of superluminous supernovae \citep{eftekhari_2019_apjl}, long gamma-ray bursts \citep{marcote_2019_apjl}, or the population of `wandering' black holes in dwarf galaxies \citep{reines_2020_apj} --- some of which may actually be related to repeating FRBs as opposed to AGN-like activity \citep{eftekhari_2020_apj}.}

With \edit{close to two dozen} precision localisations now in hand, FRB host galaxy associations are becoming relatively routine.  However, our ability to associate the most distant FRBs to host galaxies is limited by the precision of the localisations themselves and the \edit{resulting} chance coincidence probability.  For example, for one particularly distant FRB, \citet{law_2020_apj} are only able to identify two potential host galaxies, \edit{perhaps neither of which is the true host}, despite the high precision localisation of this source.

\subsection{Seeing them in fine detail --- in time, frequency and polarisation}\label{sec:2_3}

The \edit{acquisition} of full polarisation, \edit{voltage} data for FRBs is becoming increasingly common.  \edit{Voltage} data provide excellent flexibility in exploring a range of time and frequency scales, and also allow for coherent dedispersion to be applied (see \S~4.1.2 of PHL19 for a review).  Both repeating and non-repeating FRBs have previously shown temporal structure as short as 20--30\,$\upmu$s \citep{farah_2018_mnras,michilli_2018_natur}.  \citet{cho_2020_apjl} investigate a one-off FRB in detail and find complicated polarimetric properties in four narrow sub-bursts.  They also find variable apparent rotation measure (RM) and DM between these sub-bursts.  \citet{day_2020_mnras} investigate the polarimetric properties of five ASKAP discoveries and find that the one repeating source in their sample is similar to other known repeaters (about 100\% linearly polarized, with flat \edit{polarisation} position angle and no evidence of circular polarisation), while the one-off sources are more heterogeneous in their polarimetric properties.

\edit{In contrast, \citet{luo_2020_natur} show polarisation angle swings in some of the bursts they observed from the repeater FRB~20180301A.  These are reminiscent of the polarisation angle swings seen from pulsars, and suggest an origin in a rotating magnetosphere --- though other interpretations remain possible.  Significant circular polarisation has now also been detected in some bursts from the repeater FRB~20201124A \citep{hilmarsson_2021_mnras,kumar_2021_arxiv,xu_2021_arxiv}.  While some of the observed circular polarisation may have been converted from linear via propagation effects in the local medium, it appears that most of the signal must be related to the emission mechanism itself.}

\edit{Other} recent studies of repeaters have shown that their brightness varies on a wide range of timescales, from $<100$\,ns up to tens of milliseconds \citep{nimmo_2021_natas,nimmo_2021_arxiv_arxiv210511446,majid_2021_apjl}.  \edit{Indeed, while \citet{nimmo_2021_natas,nimmo_2021_arxiv_arxiv210511446} find nano- and microsecond structures in some repeater bursts, they also emphasise that not {\it all} bursts display sub-millisecond temporal structure (or that such structure must be very closely spaced, as opposed to isolated microsecond shots).} \edit{At typical observing frequencies of $\sim1.4$\,GHz} these studies are limited by the Milky Way foreground, which imparts at least tens of nanoseconds of temporal broadening due to scattering for most lines of sight.  \edit{Characterisation of (sub-)nanosecond FRB structure thus requires follow-up at higher radio frequencies ($\gtrsim3$\,GHz).}

The narrow-band nature of repeaters \citep{gourdji_2019_apjl} has inspired searches that are better tuned to picking out FRBs in both the time and frequency dimensions.  \citet{kumar_2021_mnras} used the Parkes Ultra-Wideband Low (UWL) receiver to discover a burst from FRB~20190711A whose spectral extent within the 0.7--4.0\,GHz band was only 65\,MHz.  \edit{Simultaneous} multi-band, multi-telescope observations have also demonstrated such behaviour \citep{chawla_2020_apjl,majid_2020_apjl,pearlman_2020_apjl,pleunis_2021_apjl,pastormarazuela_2021_natur}. \citet{levkov_2020_arxiv} investigate periodic spectral structure in high-frequency data from \aofrb\ and discuss possible interpretations.  \edit{FRBs also show the expected scintillation from the Milky Way ISM \citep{hessels_2019_apjl,marcote_2020_natur,schoen_2021_rnaas}.}

Targeted observations of repeating FRBs have also expanded the range of radio frequencies over which FRBs have been detected, now from 110\,MHz up to 8\,GHz \citep{pleunis_2021_apjl,pastormarazuela_2021_natur,chawla_2020_apjl,pilia_2020_apjl,majid_2020_apjl,michilli_2018_natur,gajjar_2018_apj}.  While the LOFAR detection of some bursts from FRB~20180916B shows 150-ms scattering at 110\,MHz, the brightness of at least one burst at these low frequencies suggests that some FRBs may also be detectable below 100\,MHz.  \edit{Similarly, FRB emission is likely also to be detectable at frequencies $>8$\,GHz, but such searches are challenging because the small dispersive delay makes it more difficult to differentiate between astrophysical emission and human-made radio interference.}

\subsection{Seeing them change in time}\label{sec:2_4}

The detection of multiple bursts from repeaters allows properties like activity level, spectrum, DM, and RM to be measured as a function of time. The periodic activity level of the repeaters FRB~20180916B \citep[$P_\mathrm{activity} = 16.33 \pm 0.12\,\mathrm{days}$;][]{chime_2020_natur_587,pleunis_2021_apjl} and \aofrb\ \citep[$P_{\rm activity} \sim 160$\,days;][]{rajwade_2020_mnras,cruces_2021_mnras} is potentially an important insight into the nature of these sources (see \S~\ref{sec:3_3}). The known gamma-ray binaries in the Milky Way \citep{dubus_2013_aarv} present a potential Galactic analogue; the recent discovery of a magnetar in such a system \citep{yoneda_2020_phrvl} adds weight to this possible link. Similarly, the 216.8-ms periodicity observed between sub-bursts in the one-off FRB~20191221A strongly suggests a neutron star link \citep{chime_2021_arxiv_arxiv210708463}.  At the same time, it is curious that only one out of hundreds of FRB sources has clearly shown this behaviour.

\begin{figure}
    \centerline{
    \includegraphics[width=0.9\columnwidth]{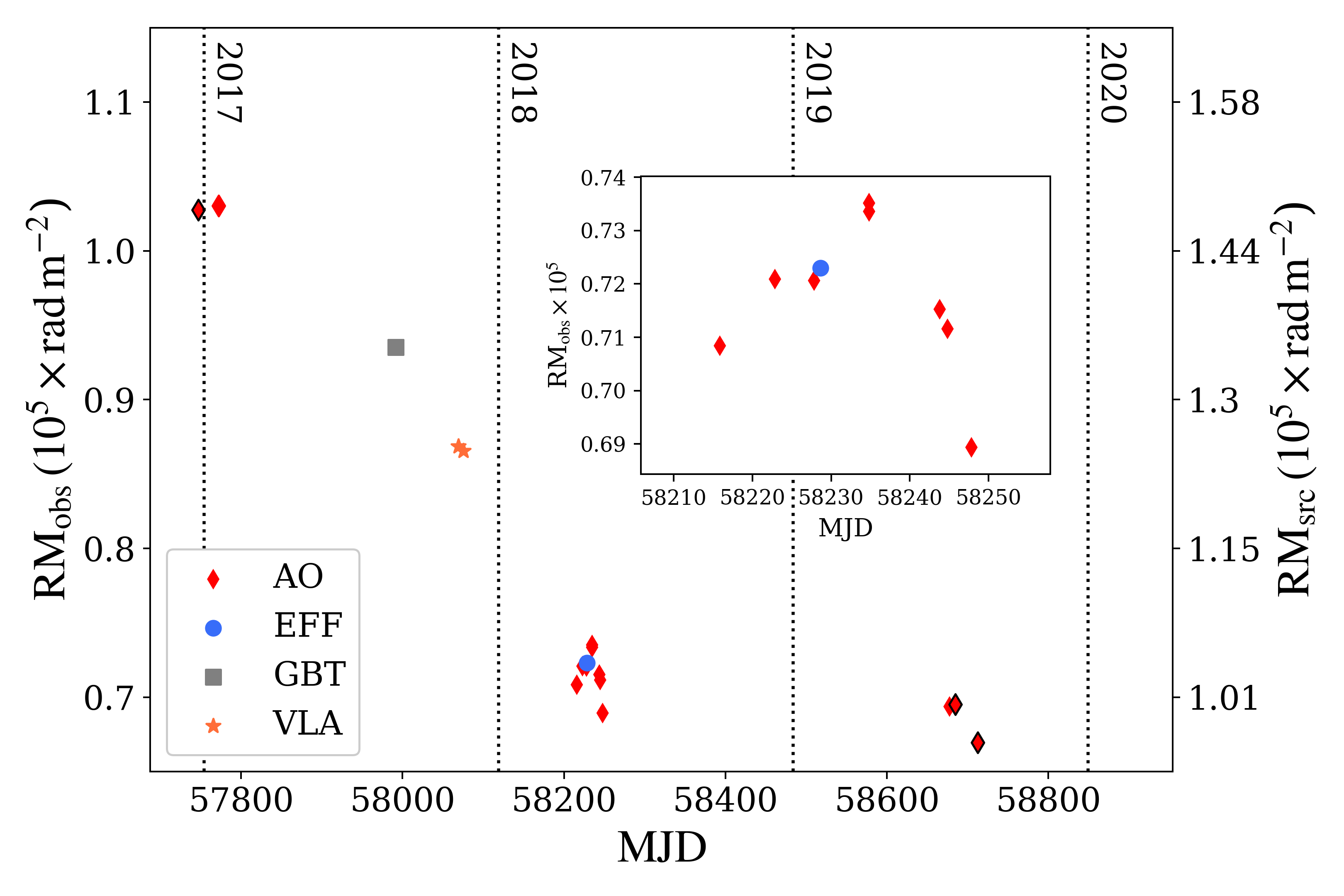}}
    \caption{\edit{Observed (left axis) and source frame (right axis) rotation measures of bursts from \aofrb, as a function of time.  The source has shown both a general decrease of $\sim35000$\,rad~m$^{-2}$, over the course of 3 years, as well as smaller 100--1000\,rad~m$^{-2}$ variations on timescales of days to weeks (see, e.g., inset).  Figure from \citet{hilmarsson_2021_apjl}.}}
    \label{fig:rm_variations}
\end{figure}

DM and RM variations have been seen from the original repeater, \aofrb\ \citep{hessels_2019_apjl,michilli_2018_natur}.  In an 8-yr span (2012--2020), the DM has increased from $557$\,cm$^{-3}$\,pc to $564$\,cm$^{-3}$\,pc \citep{spitler_2016_natur,hessels_2019_apjl,hilmarsson_2021_apjl,li_2021_natur}.  This increase is most likely associated with the local burst environment.  Unfortunately, the local DM contribution is not well known, but it is very likely $< 100$\,cm$^{-3}$\,pc, meaning that the observed fractional DM change is enormous compared to what is typically seen from Galactic pulsars (where $\Delta$DM is typically $\sim 10^{-3}$\,cm$^{-3}$\,pc\,yr$^{-1}$). In contrast,  From 2017--2020, the absolute RM has decreased from $-103,000$\,rad\,m$^{-2}$ to $-67,000$\,rad\,m$^{-2}$ \citep[][note that in the source frame this corresponds to $-148,000$\,rad\,m$^{-2}$ to $-96,000$\,rad\,m$^{-2}$]{hilmarsson_2021_apjl}. \edit{\aofrb\ has also shown RM variations of 100s to 1000s of rad\,m$^{-2}$ on timescales of days to weeks \citep{michilli_2018_natur,hilmarsson_2021_apjl}.  Other repeaters have also shown such short-timescale RM variations at a similar magnitude \citep{luo_2020_natur,xu_2021_arxiv}, while others have been considerably more stable, showing only subtle RM variations \citep{pleunis_2021_apjl}.}  Foreground and host galaxy ISM contributions are comparatively small, and hence \edit{such changes directly map} variations in the local environment.

\subsection{Seeing them at other wavelengths}\label{sec:2_5}

Despite being, until recently, the closest-known extragalactic FRB \edit{source}, FRB~20180916B shows no evidence for prompt or persistent optical or high-energy X-ray or gamma-ray emission \citep{scholz_2020_apj,pilia_2020_apjl}.  At 3.6\,Mpc, FRB~20200120E provides the best opportunity yet for multi-wavelength detections, but no such emission has yet been found \citep[][these limits rule out giant flares that might be expected from a young hyper-active magnetar]{kirsten_2021_arxiv,mereghetti_2021_apjl}.

\begin{figure}
    \centerline{
    \includegraphics[width=0.9\columnwidth]{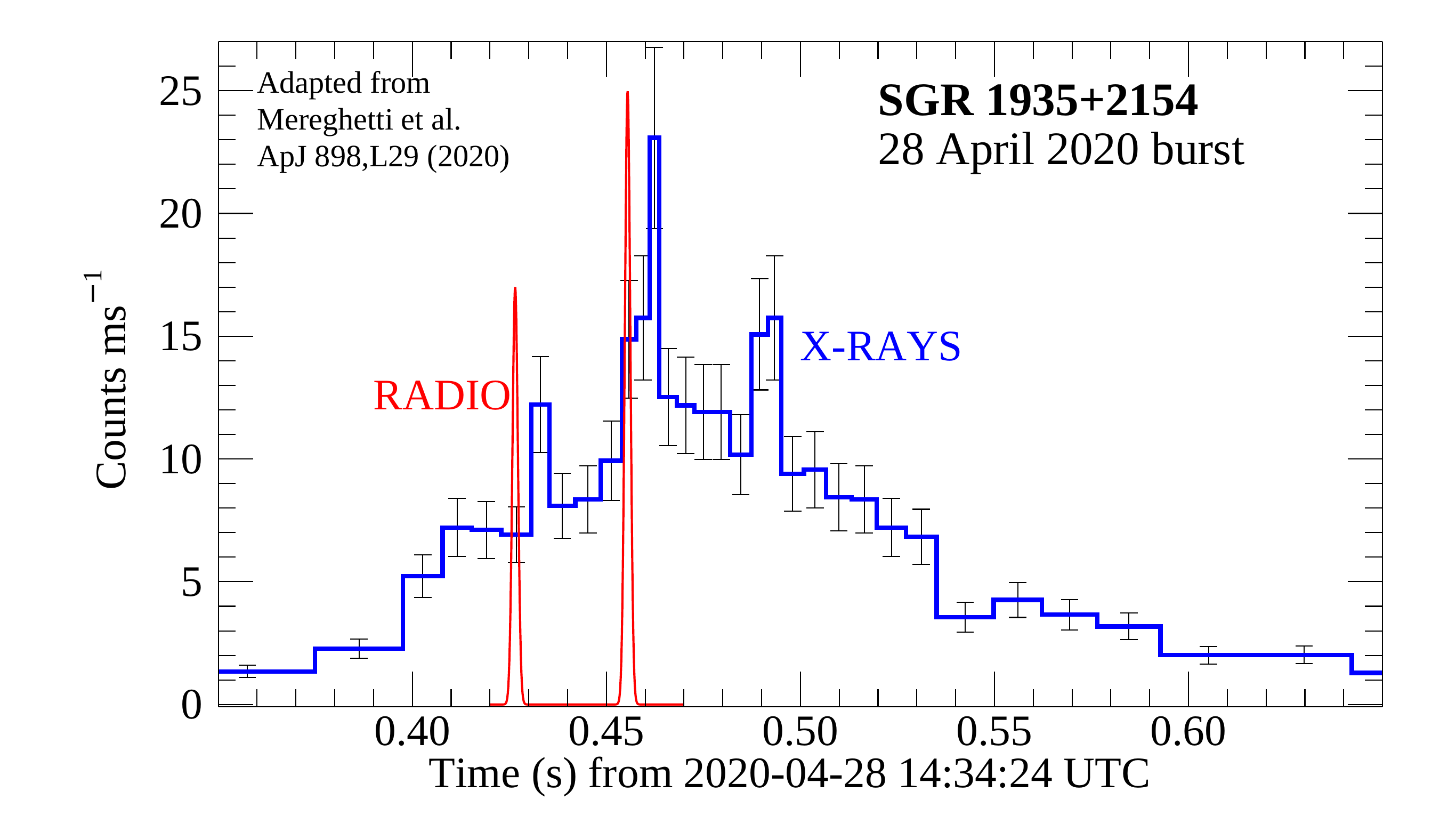}}
    \caption{\edit{Hard X-ray burst detected from SGR~1935+2154 using {\it INTEGRAL}.  Peaks 1 and 2 roughly align with the two radio peaks seen from the CHIME/FRB detection \citep{chime_2020_natur_587}, though there is a lag of a few milliseconds.  Figure courtesy S.~Mereghetti, adapted from \citet{mereghetti_2020_apjl}.}}
    \label{fig:sgr_x-ray}
\end{figure}
    
SGR~1935+2154 provides arguably the most interesting multi-wavelength information to date.  INTEGRAL observations in the hard X-ray \edit{band} (20--200\,keV) show a burst\footnote{\edit{A number of other high-energy telescopes also detected this burst.}} lasting for $\sim 0.6$\,s, and superimposed on this broad feature are three narrow ($\sim3$\,ms) peaks separated by $\sim29$\,ms.  Two of the narrow peaks match with radio bursts seen in the CHIME/FRB 400--800\,MHz and STARE2 (1.4\,GHz) bands, but with a delay of $6.5\pm1.0$\,ms from the radio bursts \citep{mereghetti_2020_apjl}.

\section{Observational constraints on theory}\label{sec:theory}
    
Prior to PHL19, the  number of known FRBs and theories for their origins grew roughly at the same pace. Now, with the much larger sample, and more well-studied sources, observations can begin to \edit{better} constrain theoretical models for FRB progenitor systems and emission. \edit{In the following subsections we discuss how measurements of the burst properties, host and local environments, and overall population statistics can constrain current models for radiation mechanisms and progenitors.}
    
\subsection{Constraining the radiation mechanism}\label{sec:3_1}

\begin{figure}
    \centering
    \subfloat[Shock models\label{fig:ShockModel}]{\includegraphics[width=0.65\textwidth]{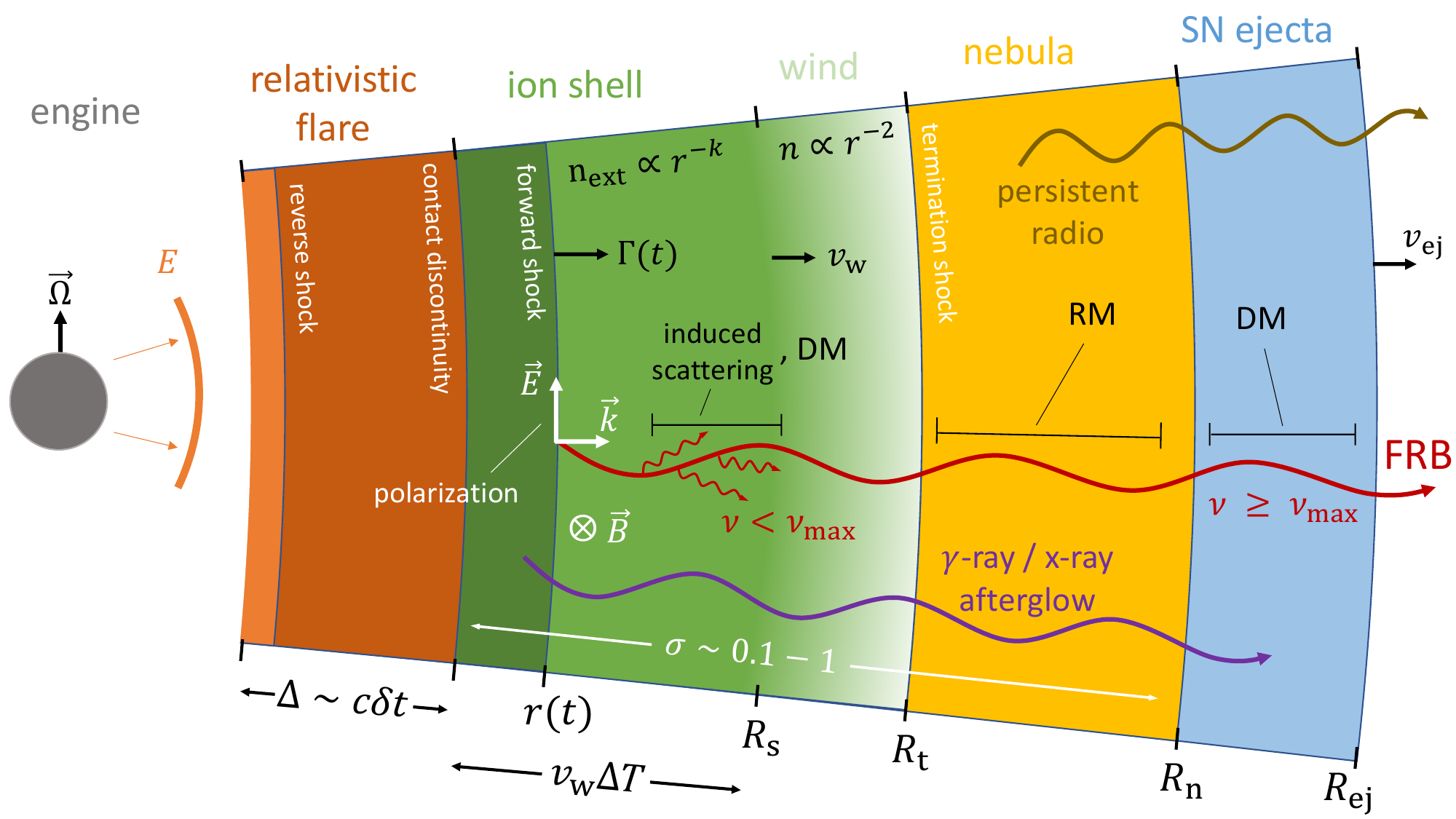}}
    \hfill
    \subfloat[Magnetospheric models\label{fig:MagnetosphereModel}]{\includegraphics[width=0.3\textwidth]{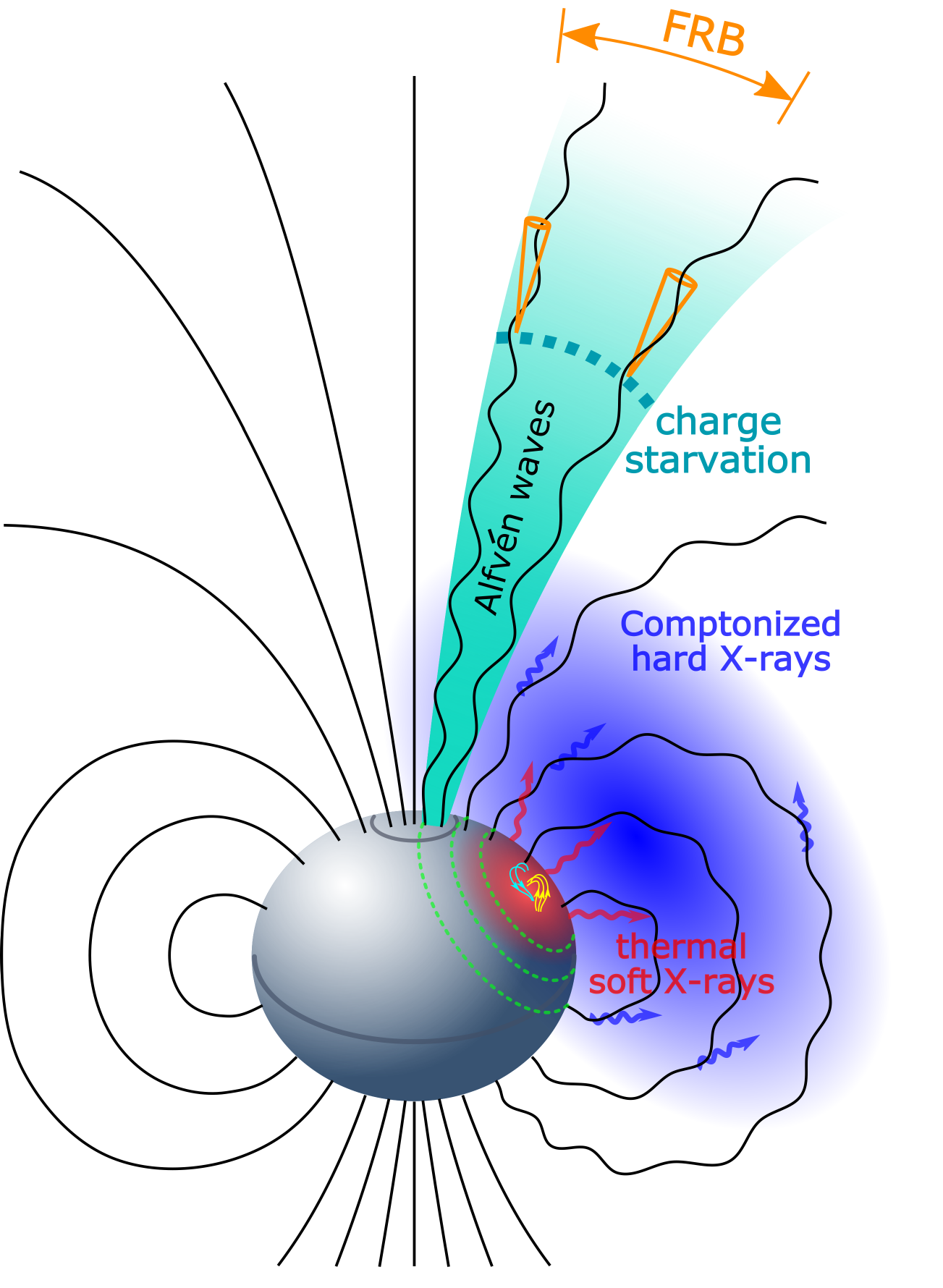}}
    \caption{\edit{Different models of FRB emission. (a) Shock model from \citet{metzger_2019_mnras} where the FRB is produced at large ($10^{10}$\,cm) radii from the compact central engine (e.g., a magnetar or black hole). (b) Magnetospheric model from \citet{lu_2020_mnras} where the FRB is produced in the neutron star magnetosphere.}}
    \label{fig:EmissionRegion}
\end{figure}

While FRB emission is widely accepted to be generated in a coherent process, the exact radiation mechanism is the subject of debate. Most models center around an energetic neutron star or magnetar progenitor but place the location of emission at different radial distances from the surface of the star \edit{(Fig.~\ref{fig:EmissionRegion})}. In some models the FRB is generated by reconnection in the neutron star magnetosphere \citep{lyutikov_2020_arxiv,lyutikov_2021_apj} or via coherent curvature radiation near the neutron star surface \citep{kumar_2017_mnras}; in both cases, the site of emission is at a radial distance of $\lesssim 10^{4}$\,km. In other models the FRB is generated via synchrotron maser emission in the forward shock of a propagating flare from the magnetar as it collides with the surrounding medium at radial distances of $\gtrsim 10^{5}$\,km \edit{\citep{metzger_2019_mnras}}.

It is difficult to differentiate between these models based on most observed properties of FRBs, as both flavors of model have been able to explain most observations to-date with relative comfort. However, recent observations of microstructure within FRB pulses, on timescales $\lesssim 10~\upmu$s (see \S~\ref{sec:2_3}) may rule out mechanisms where FRBs are generated at large radii from the central neutron star \citep{nimmo_2021_natas,nimmo_2021_arxiv_arxiv210511446,majid_2021_apjl}. The minimum timescale for emission is constrained as
\begin{equation}
    \Delta t = R \zeta^2 / 2 c \Gamma^2,
\end{equation}
where $R$ is the radial distance from the central object, $\zeta$ is the radiative efficiency, and $\Gamma$ is the Lorentz factor \citep{beniamini_2020_mnras}. A \edit{conservative estimation of the} Lorentz factor $\Gamma \sim 10$ \edit{\cite{kumar_2017_mnras,waxman_2017_apj}}, $\zeta \ll 1$, and $R \sim 10^{5}$\,km give $\Delta t$ $\sim 10~\upmu$s. Variability on shorter timescales than this would be difficult to explain for a model where the FRB is produced at large radii from the compact object. More observations of sub-microsecond structure in FRBs could differentiate between radiation mechanism models.

Similarly, the observed \edit{sub-pulse} periodicity \edit{of} FRB~\edit{20191221A} from the CHIME/FRB sample, \edit{and two further bursts with tentative periodicity}, provides additional support to a magnetospheric origin of FRB emission as strict periodicity between many sub-bursts is more difficult to achieve with propagating shocks \citep{chime_2021_arxiv_arxiv210708463}. Sub-burst periodicity has already been observed in a Galactic magnetar by \citet{levin_2019_mnras}. Additionally, \citet{wadiasingh_2020_apjl} have proposed that quasi-periodic oscillations of the neutron star crust could manifest as periodic separation between FRB sub-bursts on millisecond timescales. Future observations of burst trains may provide more detailed information about this phenomenon.

Further constraints on the radiative mechanism may be possible with the detection of individual FRB pulses over a decade or more in radio frequency. While bursts from FRB~20180916B have been seen from 110\,MHz to 1.7\,GHz \citep{pleunis_2021_apjl,nimmo_2021_natas}, no bursts have been seen \edit{simultaneously across a wide frequency range} despite commensal searches at low and high radio frequencies \citep{chawla_2020_apjl,pearlman_2020_apjl,pleunis_2021_apjl,pastormarazuela_2021_natur}.  \edit{Dual-band (2.4\,GHz and 8.3\,GHz) observations of \aofrb\ have also found the bursts to be narrow band \citep{majid_2020_apjl}.} The \edit{total emitted energy of FRBs, or rate of observed bursts above a nominal fluence threshold} at low and high frequencies would provide a further useful discriminator between emission models. \edit{Most dual-frequency studies of FRBs to-date have been conducted on repeaters, but non-repeaters may ultimately prove more promising as the one-off population appears more broadband than the CHIME/FRB repeater sample \cite{pleunis_2021_apj}}.  

\subsection{Constraining the progenitor}\label{sec:3_2}

Even more \edit{so} than radiation mechanisms, the progenitor engines of FRBs are broadly debated in the literature. Most unique entries in the FRB Theory Catalogue\footnote{\url{https://frbtheorycat.org/index.php}} focus on the FRB progenitor \citep{platts_2019_phr}. These theories range broadly to encompass neutron star emission (both from isolated and binary systems), winds in ultra-luminous X-ray sources (ULXs), compact object mergers (neutron stars, white dwarfs, black holes), active galactic nuclei (AGN), and superconducting cosmic strings. See \S~9 of PHL19 for more details on these models.

The strongest constraints on progenitors still come from the observed repetition of some FRBs \citep{spitler_2016_natur}, ruling out cataclysmic engines for these sources. Whether all FRBs repeat remains an open question \citep{caleb_2019_mnras_484,chime_2021_arxiv_arxiv210604352}, but only a non-destructive process capable of producing bursts separated by seconds, minutes, and even years can explain repeaters. Repeating bursts also place lower limits on the total energy budget of the progenitor, which must be capable of powering all observed pulses, and potentially also additional pulses if emission is beamed and not always aligned towards the observer on Earth. The strongest constraints come from the longest-lived active source seen to date, \aofrb. \citet{margalit_2020_mnras} estimate a total energy budget of \edit{$10^{47}$--$10^{49}$\,erg} for this source which they note is consistent with the magnetic energy reservoir of a magnetar; however, this number may change with the large samples of pulses from this source observed by the FAST and Arecibo telescopes \citep{li_2021_natur,aggarwal_2021_apj,hewitt_2021_arxiv}. More active repeaters, sources with larger burst energies, and longer monitoring of currently active repeaters could place further constraints on this reservoir through future observations, as could observations of fading or quiescence in currently active sources.

Now that a small subset of FRBs have been precisely localised to host galaxies (see \S~\ref{sec:observations}), the properties of the FRB hosts can also be compared to other known populations of astrophysical transients and populations. Thus far, FRB distributions have been found to be inconsistent with long gamma-ray burst (GRB) offsets \citep{heintz_2020_apj, mannings_2021_apj}, and host galaxy stellar mass and star formation rates are for the most part inconsistent with channels such as hydrogen-poor superluminous supernovae \citep{bochenek_2021_apjl}. However, the \edit{FRB} population remains consistent \edit{with being powered by} magnetars formed in core-collapse supernovae (CCSNe) --- with the notable exception of FRB~20200120E, which was localised to a globular cluster by \citet{kirsten_2021_arxiv}. Further host galaxy studies may also compare the FRB source positions and host galaxy metallicities to those of ULXs, which tend to reside near (but not necessarily within) young stellar clusters in low metallicity hosts \citep{sridhar_2021_apj}.

In the future, a direct association of an FRB event and another astrophysical transient would provide a conclusive link between their progenitors. Further insights into the FRB progenitor(s) could come from a larger sample of host galaxies, which might reveal sub-classes of FRBs associated with specific host \edit{types}, or conclusive evidence that the FRB progenitor population evolves over cosmological time --- \edit{e.g., following the} star formation rate or other host galaxy properties (metallicity, galaxy mass, etc.).

\subsection{Constraining the system}\label{sec:3_3}

\begin{figure}
    \centering
    \includegraphics[width=0.9\textwidth]{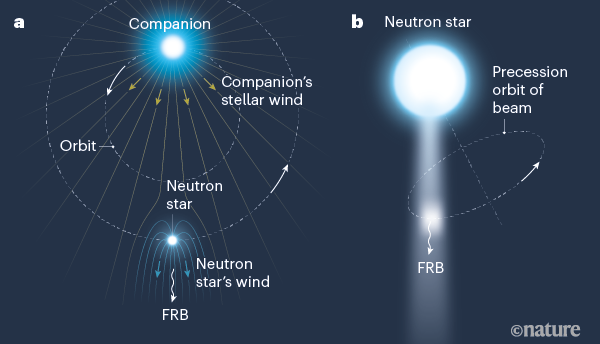}
    \caption{A diagram illustrating two systems that could produce periodic emission from FRBs, including (a) a neutron star and a binary companion, and (b) a precessing isolated neutron star. Figure from \citet{zhang_2020_natur}.}
    \label{fig:zhang_periodicity}
\end{figure}

Since the discovery of periodic activity from FRB~20180916B, and \edit{likely} periodicity in \aofrb, more theories have focused on binary progenitor systems. Early cataclysmic theories also suggested mergers of compact objects; however, theories of persistent binary systems have risen in popularity. The periodicity could be induced by binary motion, rotation, or precession \citep[see Fig.~\ref{fig:zhang_periodicity}, and also][]{levin_2020_apjl}.  In the case of a binary model, it could be that interaction between two orbiting bodies is a prerequisite for producing bursts \citep{zhang_2017_apjl}, though it has also been proposed that the modulation in activity level is due to absorption by intra-binary plasma \citep{lyutikov_2020_apjl}.  It has been shown that binary interaction dominates the evolution of massive stars \citep{sana_2012_sci}, and if the burst engine of FRB~20180916B and \aofrb is indeed a young magnetar then it would not be surprising to find such a source in an eccentric orbit with a massive (e.g. OB) stellar companion.

Constraints on the configuration of a binary system producing FRBs can be placed in a number of ways. The observation of wider and later peaks in the activity window at lower radio frequencies \citep{pleunis_2021_apjl,pastormarazuela_2021_natur} constrains simple binary wind models such that the periodicity might not arise from occultation of the emission by a wind.  \citet{wada_2021_apj} argue that these constraints can be accommodated in a binary wind model. Some binary models such as \citet{sridhar_2021_apj} predict that FRB emission will shut off on timescales of years to decades due to the ULX binary becoming too tight. This model also predicts a luminous red nova in the years after the FRB emission shuts off, when the two stars merge.

Longer baselines of observations for highly active repeating FRBs may allow for FRB timing, either through periodic pulse arrival times (not yet observed for any repeating FRB), or evolution of periodic activity windows \edit{like those seen} for FRB~20180916B. Long-term timing of these systems could shed more light on the nature of a binary system, analogous to \edit{pulsar timing experiments} \citep{lorimer_2012_hpa}. Additionally, precise localisation of nearby repeaters could enable the identification of an optical stellar companion through deep imaging. In this case, a direct measurement of the binary companion and nature of the system would be possible; however, this would only be feasible for the very nearest FRBs. Deeper imaging of the globular cluster containing FRB~20200120E (see \S~\ref{sec:2_2}) could provide \edit{more information about the dynamic interactions and environment of this FRB source.}

\subsection{Constraining the environment}\label{sec:3_4}

Current FRB observations have already put direct constraints on the local environments of known repeating and one-off FRBs. Even for poorly localised sources, the RM excess of an FRB relative to the Galactic foreground can indicate a clean environment in the case of no excess, or a highly magnetised and highly ionised local environment in the case of high RM values. The canonical example is \aofrb\ (see \S~5.4 of PHL19), however more than 10 FRBs have now been identified with significant RM excess. \edit{These sources indicate} that substantial magneto-ionic material local to the source is relatively common for FRBs, even for those that have not been seen to repeat.

In the case of \aofrb, the presence of a \edit{compact} persistent radio source co-located with the FRB has also informed studies of the local environment \citep{chatterjee_2017_natur,marcote_2017_apjl,tendulkar_2017_apjl,bassa_2017_apjl}. The extremely high RM \edit{of the bursts} has been attributed to this local \edit{radio} source, hypothesised to be a nebula or \edit{accreting massive black hole} (see \S~5.4 of PHL19). Persistent radio sources associated with other FRBs may indicate similarly extreme environments. \edit{Only two other FRB sources have been found to be coincident with persistent radio sources, the repeaters FRB~20201124A and FRB~20190520B.} \edit{In the case of FRB~20201124A, \citet{fong_2021_apjl} and \citet{ravi_2021_arxiv} attribute the radio emission to ongoing star formation, likely in the circumnuclear region, later confirmed by \citet{piro_2021_aa}}. Observations of an FRB's \edit{direct} environment require precise localisation, preferably to sub-arcsecond precision so as to identify the local environment within the host as has been done for several FRBs from ASKAP \citep{bhandari_2020_apjl_895}, as well as for \edit{FRBs~20180916B, 20200120E, and 20201124A \citep{tendulkar_2021_apjl,kirsten_2021_arxiv,nimmo_2021_arxiv_arxiv211101600}}. 

In the absence of localisation, radio observations of FRB pulses themselves still have incredible power to probe the local FRB environment. Observations at low radio frequencies ($\sim100$\,MHz) can put direct upper limits on the density of the local environment by constraining absorption in the local ionised medium \citep{pleunis_2021_apjl,pastormarazuela_2021_natur}. 

Long-term monitoring of variations in DM, RM and scattering for repeating FRBs can also place constraints on the local magneto-ionic material in the FRB progenitor environment (see \S~\ref{sec:2_4}). Significant variations in these parameters correspond to the transverse movement or dissipation of filaments, perhaps in a nebula or ionised wind. The classic example is \aofrb\ \citep{hilmarsson_2021_apjl}, although even smaller order variations are constraining \citep{luo_2020_natur,pleunis_2021_apjl}. 

\subsection{Constraining the population}\label{sec:3_5}

With a sample of hundreds to thousands of FRBs, we can also constrain the population properties of the sources and their distribution over cosmic distance. Thus far only the CHIME/FRB Catalogue sample has been of sufficient size for the purpose of studying distribution in DM, duration and scattering \edit{\citep{Chawla_2021_arXiv}}, but the imprecise localisations and redshifts of the sample make studies of the intrinsic spatial distribution challenging \citep{chime_2021_arxiv_arxiv210604352}.

Until a large sample of localised FRBs is available, population synthesis studies can model the FRB population and its distribution under a variety of assumptions such as if FRBs were to trace the star formation rate (SFR) history of the Universe. Population synthesis can also attempt to convolve these modeled populations with instrumental effects from telescopes, to estimate what the observed population distributions would look like over DM, pulse width, scattering, and fluence \citep{gardenier_2019_aa,james_2022_mnras_1}. 

\begin{figure}
    \centering
    \includegraphics[width=0.9\linewidth]{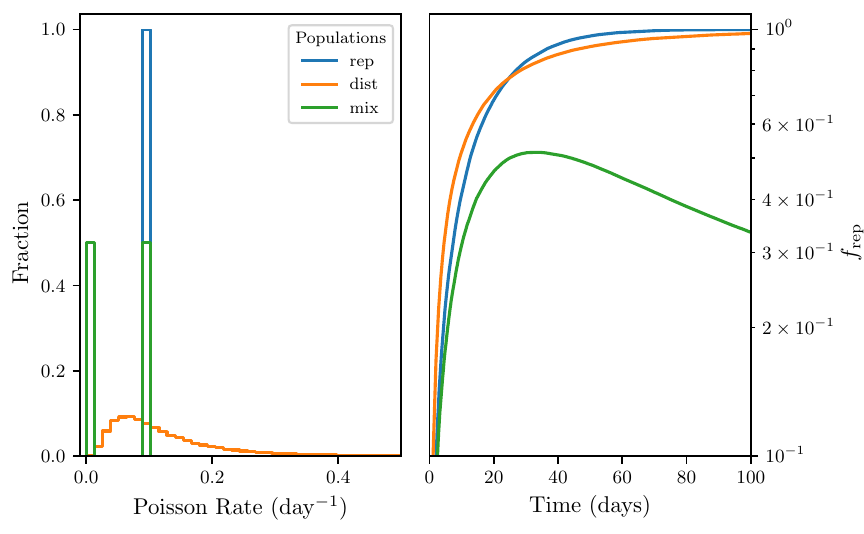}
    \caption{Fraction of FRBs observed as repeaters (right) using the \texttt{frbpoppy} population synthesis code for various assumed intrinsic populations (left). For assumptions of a distribution of repeat rates, a 100\% repeating population, and a mix of repeating and non-repeating FRBs, the fractional number of FRBs detected by telescopes as repeaters as a function of time will change and eventually asymptote or turn over. Modified from \citet{gardenier_2021_aa}, courtesy of J. van Leeuwen.}
    \label{fig:frbpoppy}
\end{figure}

A recent population synthesis on a small sample of Parkes and ASKAP bursts suggests that the population evolves with SFR \citep{james_2022_mnras_2}. \edit{In their study, however, \citet{zhang_2021_mnras} cannot distinguish between a SFR model from one in which the population tracks compact binary mergers.}  Future studies with a larger and more precise sample could be used in turn to probe cosmic star formation. Additionally, population synthesis efforts that take into account properties of repeaters and apparent non-repeaters provide predictive models, for example: the fraction of sources seen to repeat as a function of time \citep{gardenier_2021_aa}. If repeating sources represent a separate sub-class of FRB progenitors, then the repeater fraction may plateau or decrease over time as the total number of FRBs increases (Fig.~\ref{fig:frbpoppy}).

Thus far, population synthesis methods have had limited ability to distinguish between progenitor models of FRBs. \citet{james_2022_mnras_2} have noted that their models are consistent with a population arising from magnetar bursts. However, conclusively ruling out progenitor classes based on population synthesis will likely have to wait until the observed sample being compared to models is of sufficient size to be robust, requiring hundreds and perhaps thousands of FRBs.

\subsection{What can we learn from SGR~1935+2154 and other \edit{known} magnetars?}\label{sec:3_6}

The recent discovery of \edit{a} MJy-ms radio burst from \edit{the} Galactic magnetar SGR~1935+2154 (\S~\ref{sec:2_2}) has given us potential new ways to explore the FRB---magnetar connection. Observations suggest that a bright SGR flare (such as might be seen from another galaxy as an FRB) might also be followed by several bursts, orders of magnitude weaker in fluence, in the days and weeks following the initial flare.  \edit{Such bursts were seen in the case of SGR~1935+2154} \citep{kirsten_2021_natas}. Follow-up of nearby, bright FRBs of fluence $\gtrsim$10 Jy-ms with sensitive telescopes like FAST, GBT, or (in the future) CHORD or the SKA, could detect these lower fluence aftershocks. Hard X-ray flares coincident with (or slightly lagging) radio-detected FRBs might also prove telling \edit{(See Fig.~\ref{fig:sgr_x-ray})}. Such follow-up is already ongoing with the nearest reported FRB~20200120E in M81 \citep{bhardwaj_2021_apjl_910}. Similarly, searches for FRBs with spectra similar to that of the radio burst from SGR~1935+2154, seen by CHIME/FRB to be double-peaked and perhaps upward drifting in frequency, could lead to further insights.

Recent studies of the short gamma-ray burst population suggest that a handful of these sources with particularly \edit{hard} X-ray spectra post-burst are actually from extragalactic magnetar progenitors\edit{, and may be extragalactic versions of the `giant flares' that have been seen from some Galactic magnetars} \citep{svinkin_2021_natur,roberts_2021_natur}. These results, while as yet unrelated to FRBs, provide direct evidence for highly energetic, observable behaviour from distant magnetars. Searches for similar X-ray emission following radio bursts might also provide insight into any relationship between FRBs and magnetars.

Precise localisation of a larger number of FRBs will also help in this area, as milliarcsecond precision can help identify not only the source location in the host galaxy but also potentially associate it with a star forming region or other interesting area such as a globular cluster or H\textsc{ii} region \edit{when radio interferometric localization is combined with deep optical imaging}. In the most nearby cases, it may even be possible to find FRBs co-located with a bright companion star through sensitive imaging with {\it Hubble}, {\it James Webb Space Telescope}, or ground-based optical instruments \edit{using adaptive optics}.

Ultimately, if an FRB---magnetar connection exists, it may be necessary to find a handful of `bridging sources' that bring together traditional FRB observations (radio bursts, repetition, polarisation) with recent developments in magnetar science (X-ray spectrum, flaring behaviour).

\subsection{What questions can we answer with future observations?}\label{sec:3_7}

At present, FRB observations constrain theoretical models, but no \edit{`smoking guns' exist to conclusively support specific progenitor theories}. As the FRB population grows and observations of the larger FRB phenomenon emerge, the discovery of a few main properties may prove to be particularly impactful, \edit{particularly} related to each of the topics \edit{addressed in this section}. 
\begin{itemize}

\item \textit{Emission physics}: What is the \edit{shortest-timescale} structure in FRB bursts? Do \edit{many} repeating FRBs have observable position angle (PA) swings that vary, either over a single burst or over an entire activity window?  \edit{How do the polarimetric properties change with frequency, and is this due to intrinsic or extrinsic reasons?} 

\item \textit{Progenitors}: Does the FRB population evolve with star formation rate? 

\item \textit{Progenitor systems}: \edit{Do all repeating FRBs vary in their activity in a periodic way?} Do different repeat activity windows correspond to different binary system configurations? 

\item \textit{Environment}: Are high RMs associated with persistent radio sources? Do \edit{the} RMs of all repeaters vary over time?

\item \textit{FRB---magnetar connection}: Are FRB hosts and FRB locations in host galaxies consistent with magnetar distributions? Do \edit{other} Galactic magnetars\edit{, besides SGR~1935+2154, also} exhibit FRB-like emission?
 
\end{itemize}

\section{FRBs as astrophysical and cosmological probes}
\label{sec:probes}

Even though we do not currently fully understand the source populations and mechanisms for FRBs, as is the case for pulsars, we can use them as \edit{powerful} astrophysical probes.
We now briefly review some of \edit{their} key uses, and look ahead to what might be possible in the future. Further details can be found in \edit{an} extensive review on this topic by \citet{bhandari_2021_univ}. A recurring theme in the items below, and an area of intense ongoing study, is
the DM distribution of FRBs (after accounting for observational biases). In particular, we would like to establish the true distribution in DM, and the prospects for probing this with current \edit{and} emerging facilities. Recent studies on
the DM distribution have been carried out by  
\citet{walker_2020_aa}, \citet{arcus_2021_mnras}
and
\citet{james_2022_mnras_1}.

\subsection{Finding the missing baryons: The Macquart Relationship}\label{sec:4_1}

The exciting potential of a population of radio bursts at cosmological distances as probes of the ionised intergalactic medium was originally predicted prior to FRBs by \citet{ginzburg_1973_natur} and later noted in the final sentences of the FRB discovery paper \citep{lorimer_2007_sci}. An important application of such a population is to carry out a census of baryons in the Universe, primarily at $z<2$ where a budget from a variety of sources only accounts for about half of the total expected. As has been subsequently developed, and described in detail in PHL19, by subtracting both the expected Milky Way contribution (${\rm DM}_{\rm MW}$) for a given line of sight and the contribution from the host galaxy (${\rm DM}_{\rm host}$), each FRB provides a measurement of the electron content from the intergalactic medium along a given line of sight, 
\begin{equation}
    {\rm DM}_{\rm IGM} = 
    {\rm DM}_{\rm obs} - 
    {\rm DM}_{\rm MW}  - \frac{{\rm DM}_{\rm host}}
    {1+z}.
    \label{eq:dmcosmic}
\end{equation}
Adopting standard cosmological models \citep[see, e.g,][]{deng_2014_apjl,zhou_2014_phrvd}, the average value of intergalactic dispersion measure
\begin{eqnarray}
\langle {\rm DM}_{\rm IGM} \rangle = \Omega_b \frac{3 H_0 c}{8 \pi G m_p} \int_0^z \frac{(1+z') f_{\rm IGM} \left[ \frac{3}{4} X_{e,H}(z') + \frac{1}{8} X_{e,He}(z') \right] }{\left[ \Omega_{\rm M} (1+z')^3 + \Omega_{\Lambda}(1+z')^{3[1+w(z')]}  \right]^{1/2}} dz',  
\label{eq:dmigm}
\end{eqnarray}
where $w$ is a dark energy equation of state parameter; \edit{and} $\Omega_b$, $\Omega_{\rm M}$ and $\Omega_{\Lambda}$ are the baryonic, matter and dark energy densities, respectively, relative to the critical density, $\rho_c = 3 c^2 H_0^2/8 \pi G$.
$f_{\rm IGM}$ is the fraction of baryons produced in the Big Bang that remain in the IGM and
$X_{e,H}$ and $X_{e,He}$ are, respectively, the ionisation fractions of hydrogen and helium. 

\begin{figure}
\includegraphics[width=\linewidth]{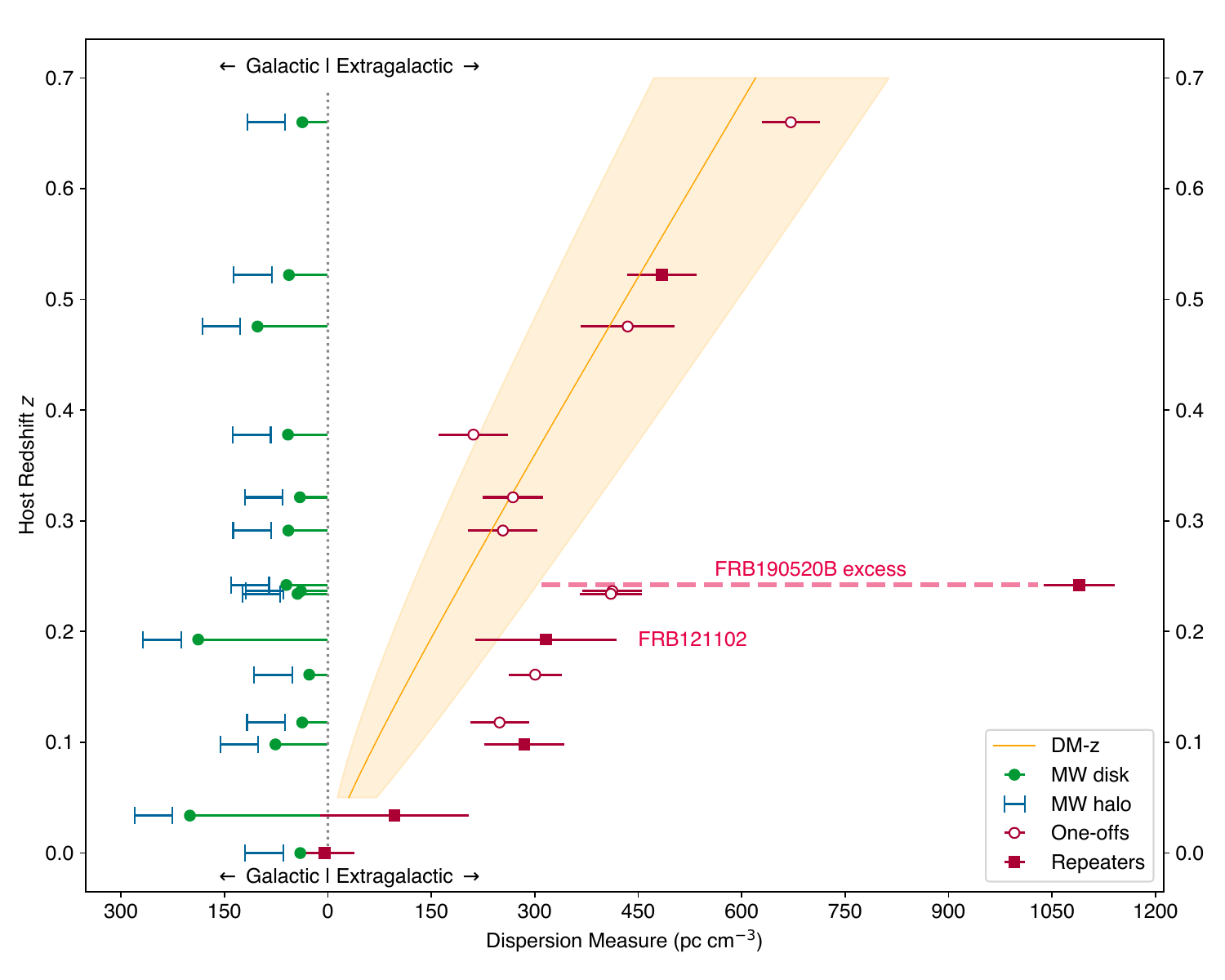}
\caption{\edit{Host redshift as a function of dispersion measure (DM) for the current sample of 14 FRBs with redshift measurements. In this split-panel display, 
  adapted from \citet{niu_2021_arxiv}, the Galactic DM contributions from the disk and the halo are shown on the left, while the remaining excess DMs are displayed in the main panel on the right. The shaded region shows the 1-$\sigma$ range around the expected relation based on Eq.~\ref{eq:dmigm}. The recent discovery of FRB~20190520B is shown as a clear outlier to this trend and indicates a significant contribution to the DM excess from the host galaxy and local environment \citep{niu_2021_arxiv}. Figure kindly provided by S.~Chatterjee.}}
\label{fig:DMz}
\end{figure}

The fundamental feature of the above expression is that $\Omega_b$ serves as the constant of proportionality between the intergalactic medium DM and the FRB redshift. Adopting nominal values of these parameters, we find that $\langle {\rm DM}_{\rm IGM} \rangle \simeq 1000 f_{\rm IGM}\,z$ for $z<1$, and $\langle {\rm DM}_{\rm IGM} \rangle \simeq 850 z$ if a typical $f_{\rm IGM}$ is adopted \citep{zhang_2018_apjl}.  This relationship is shown alongside the observed DM values and redshifts for 14 of the FRBs currently associated with host galaxies in Fig.~\ref{fig:DMz}. The figure also shows the excess DM provided by the ${\rm DM}_{\rm MW}$ and ${\rm DM}_{\rm host}$ terms in Eq.~\ref{eq:dmcosmic}. Though there are exceptions where the host and local environment contribution is large, the DMs of FRBs clearly track the expected relationship. A more detailed analysis on a smaller sample by \citet{macquart_2020_natur} was the first demonstration that FRBs probe the baryonic material in the IGM. In recognition of the pioneering work by the late J-P Macquart, who led the team that demonstrated this result, the DM--$z$ relation is now known as the Macquart Relation. 

\begin{figure}
\centering
\subfloat{\includegraphics[width=0.5\textwidth]{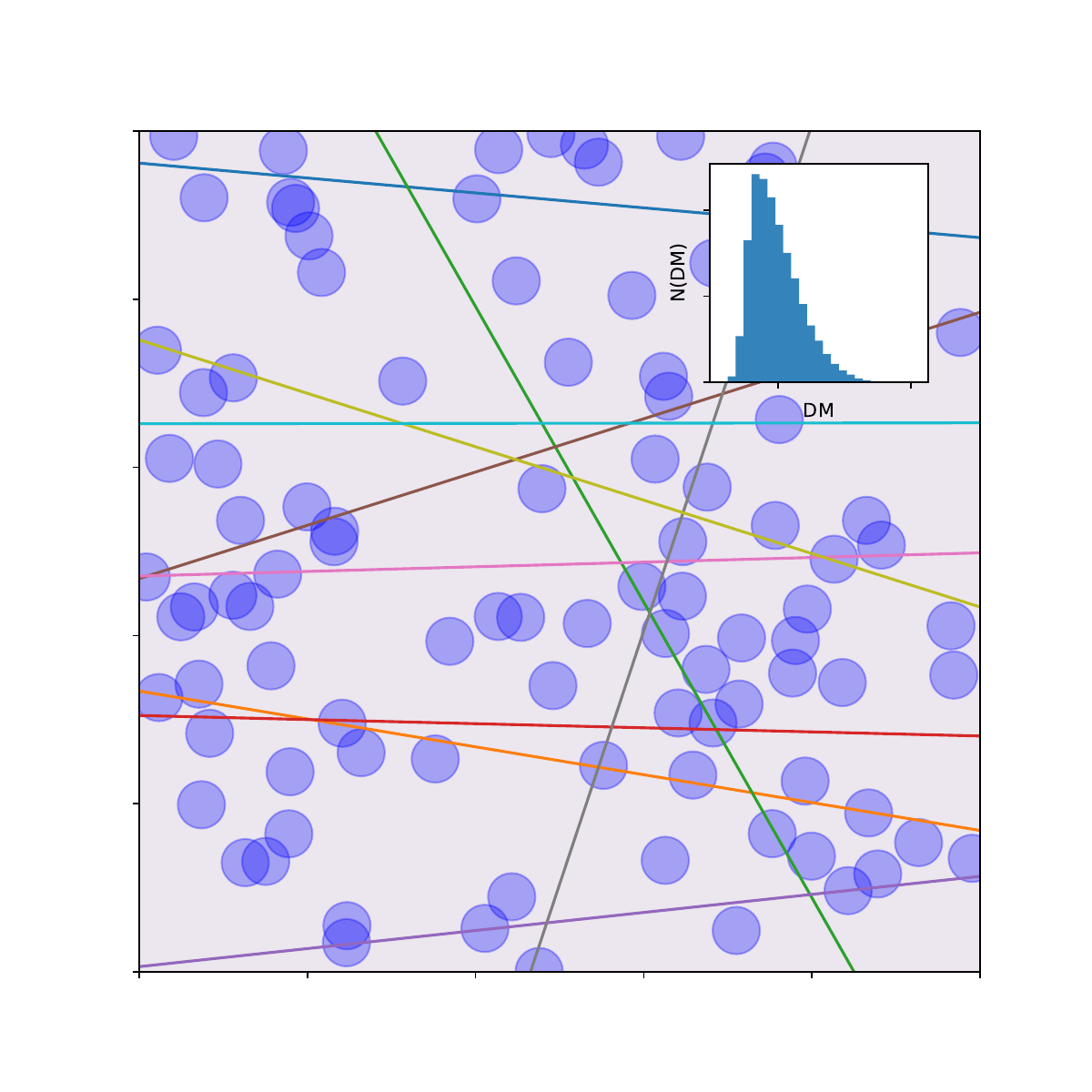}}
\hfill
\subfloat{\includegraphics[width=0.5\textwidth]{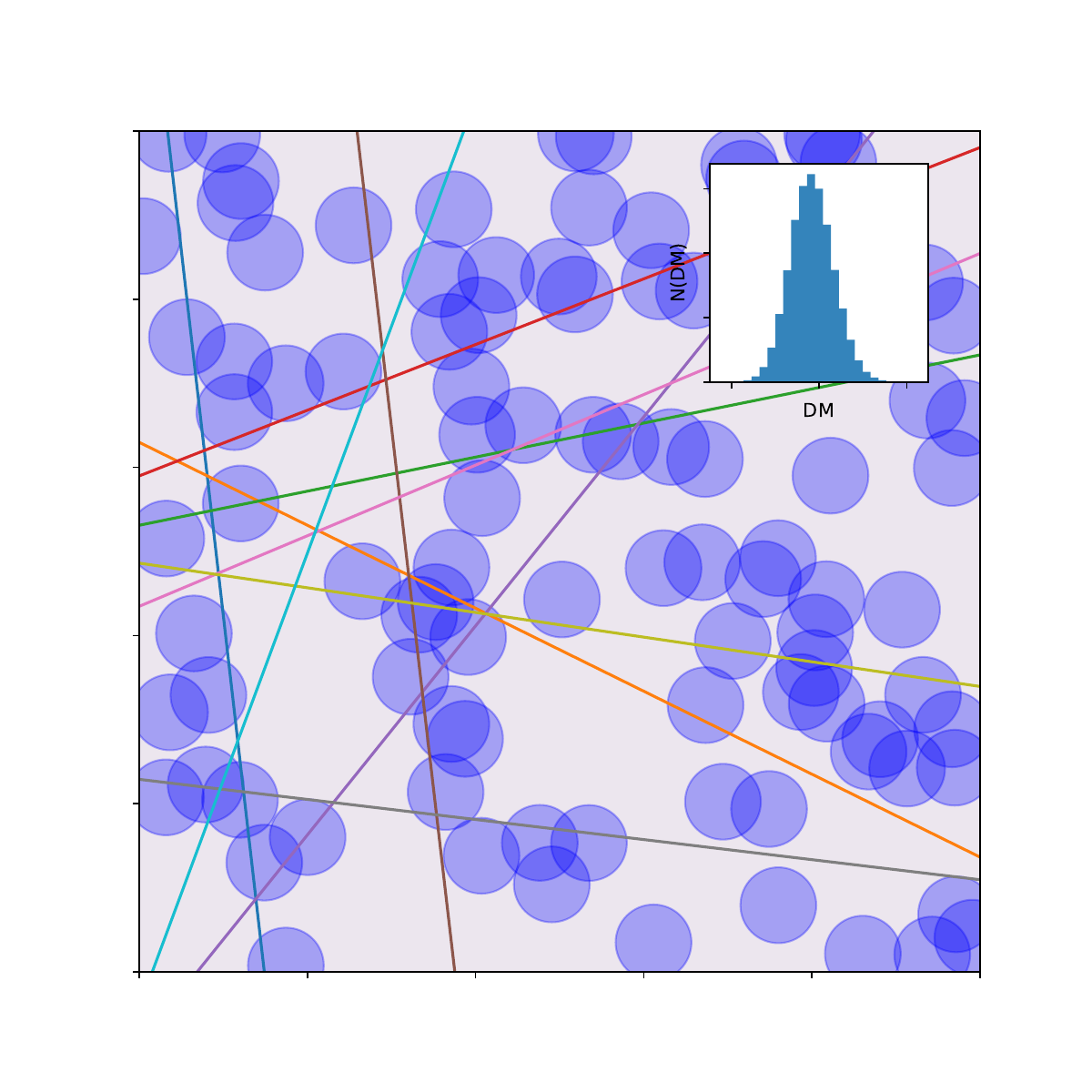}}
\caption{Schematic representation showing the dependence of the observed DM distribution on galaxy halo extent. The left panel shows a compact halo distribution, which leads to a tail in the DM distribution. The right panel shows the normal distribution, which is the result of more extended baryonic halos. The solid lines represent randomized lines of sight. Figure source: \citet{bhandari_2021_univ}.}
\label{fig:DMs}
\end{figure}

Going beyond the simple proportionality in the DM--$z$ plane, the observed variance contains important information about the baryonic halos of galaxies along the line of sight, which are governed by feedback processes from the galaxies themselves. As originally shown
by \citet{mcquinn_2014_apjl}, from a sample of around 100 well-localised and redshift-identified FRBs, it should be possible to distinguish between different baryonic matter
halo distributions in a way that is currently impossible to do by other means. For galaxies with predominantly extended baryonic halos, as shown in Fig.~\ref{fig:DMs} the DM distribution of FRBs becomes more Gaussian-like than for the case where galaxy halos are more compact, which leads to a more skewed distribution with a tail at high DMs. As demonstrated recently by \citet{walters_2019_phrvd}, a sample of around 100 FRBs with well-determined redshifts could constrain the diffuse gas fraction to the few percent level.

\edit{Another property of the missing baryons is their temperature, and how they probe the so-called `warm hot intergalactic medium' (WHIM). \citet{munoz_2018_phrvd} showed through simulations of mock FRB samples that this is possible in combination with maps of the thermal Sunyaev-Zeldovich effect. The authors forecast that a measurement of the WHIM temperature would be possible at the 10\% level with samples of $10^4$~FRBs and could definitively link the missing baryons with the WHIM.}

\subsection{Cosmic rulers}\label{sec:4_2}

The dependence of $\langle$DM$_{\rm IGM}\rangle$ on various cosmological parameters seen in Eq.~\ref{eq:dmigm} highlights the value of a well-determined set of FRBs with redshift measurements in contributing to precision cosmology. Recently, using the sample of 9 FRBs from \citet{macquart_2020_natur},
\citet{hagstotz_2021_arxiv}
use a likelihood analysis to measure $H_0=62 \pm 9$\,km~s$^{-1}$~Mpc$^{-1}$. These authors forecast that a sample of 500 FRBs with redshift determinations expected to be available in the coming decade will be able to determine $H_0$ to a precision of a few per cent or better. Such a sample would provide an independent way to help resolve the current tension between the results obtained by $H_0$ analyses of the CMB
\citep{planckcollaboration_2020_aa} and type Ia supernovae \citep{riess_2019_apj}. Similar predictions were also made prior to the
\citet{macquart_2020_natur}
sample by
\citet{wu_2020_apj}.

\subsection{Reionization history of the Universe}\label{sec:4_3}

After recombination took place in the early Universe around redshift $z=1089$, the intergalactic medium was fundamentally changed during the epoch of reionization as free electrons were liberated by ultraviolet radiation from stars and quasars in the redshift range $6<z<13$ for H\textsc{i} and He\textsc{i}. Due to the direct relationship between DM and electron density, cosmological FRBs offer new ways to probe these still poorly understood processes.

For hydrogen reionization, \citet{beniamini_2021_mnras} show that
the maximum dispersion measure for an ensemble of FRBs (DM$_{\rm max}$) can
in principle
be a powerful diagnostic of the IGM
at high redshifts. They show that,
for a sample of around 1000 FRBs with DMs above 6000\,cm$^{-3}$~pc,
a determination of DM$_{\rm max}$
with a precision of better than 10\%
would provide \edit{tighter} constraints on
the reionization history than current
measurements of CMB optical depth
using Planck
\citep[see, e.g.,][]{pagano_2020_aa}. For smaller
samples of FRBs, which are more
realisable with current and emerging
facilities, \citet{beniamini_2021_mnras}
estimate that useful constraints could
be made from 40 FRBs in $6<z<10$ given
redshift accuracies of 10\%. In the absence of redshift information, any constraints on reionization at these
epochs are strongly coupled to our (currently incomplete)
knowledge of the FRB luminosity function. 

While the Deep Synoptic Array 
\citep[aiming to become operational with 2000 6-m dishes by 2025;][]{hallinan_2019_baas} will allow for H\textsc{i} and He\textsc{i} reionization studies using high-redshift FRBs, in the near term, more progress is likely to be made for He\textsc{ii} reionization which is thought to have \edit{occurred} at redshifts as low as $z=3$
\citep{fialkov_2016_jcap}. In a recent study, \citet{caleb_2019_mnras_485} show that
1000--2000 FRBs are required to begin to probe the epoch of He\textsc{ii} reionization through their DM distribution. Further work on modeling the initial CHIME/FRB sample may well provide insights into this issue although the largest DM from that sample is 3038.06\,cm$^{-3}$~pc \citep{chime_2021_arxiv_arxiv210604352}.

\subsection{Cosmic turbulence and magnetism}\label{sec:4_4}

In addition to pulse dispersion, which probes the ionised intergalactic gas, a subset of FRBs with measurements or constraints on pulse scattering and Faraday rotation measure from polarimetry can be used to probe both the turbulent and magnetised nature of the intervening medium. As recently demonstrated by \citet{xu_2020_apjl}, the FRB sample is now beginning to be large enough to make meaningful statements on intergalactic turbulence. From a measurement of the structure function of DMs based on a sample of 112 FRBs, which probe length scales around $10^8$\,pc, it is found that the DMs attributable to the intergalactic medium are consistent with a Kolmogorov model for the turbulent medium. A similar trend is seen in the interstellar medium based on pulsar measurements \citep{armstrong_1981_natur} on sub-pc length scales. A recent analysis by \citet{zhu_2021_apj} also considers the impacts of foreground galaxies on the overall turbulence and shows that, while the IGM is still the dominant factor, intervening galaxies and their halos contribute at about the 30\% level.

\subsection{Gravitational lensing}\label{sec:4_5}

Gravitational lensing, the phenomenon whereby the light from a distant celestial object \edit{follows curved geodesics} due to the warping of spacetime by intervening objects, is now a well studied astrophysical phenomenon in samples of quasars \citep[for a comprehensive review, see e.g.,][]{blandford_1992_araa}.
While a number of discussions of the potential for gravitational lensing within FRB observations can be found in the literature (one of the earliest being by \citet{li_2014_scpma} as a diagnostic for distinguishing between progenitor models), so far
no examples of  gravitational lensing of FRB signals in the classical sense of an intervening galaxy have been found.

Recently, \citet{chen_2021_apj} have investigated \edit{gravitational} lensing in the context of double-peaked structures seen in some FRB profiles (e.g.~FRB~20130729A) which can potentially be explained in terms of a point mass (where the leading peak is always more magnified) with or without an external shear lens model. The presence of an external shear will change the relative amplitude of the two peaks. A conclusive \edit{demonstration} of \edit{gravitational} lensing, however, will require high-resolution \edit{radio} imaging.

\citet{liu_2019_phrvd} and \citet{wucknitz_2021_aa} have examined the prospects for using \edit{gravitational} lenses \edit{as cosmological probes}. Considering a potential future sample of a few dozen repeating sources that exhibit strong lensing, \edit{\citet{li_2018_natco} and} \citet{liu_2019_phrvd} show that constraints on both $H_0$ and the equation of state of dark energy are possible. \citet{wucknitz_2021_aa} discuss how measurements of time delays and their derivatives might be used to overcome limitations in conventional lensing interpretations, which typically require assumptions to be made about the unknown mass distribution of the lens. For multiply lensed sources, the images themselves might someday be utilised as an exquisitely sensitive intergalactic interferometer that could reveal structures on km scales. 

\edit{ Another novel application of lensed FRBs is their potential as dark matter probes, in particular the possible contribution from massive compact halo objects (MACHOs) which is poorly understood. 
\citet{munoz_2016_phrvl} show that,
for masses in the 20--100\,$M_{\odot}$ range,
the lensing of FRBs by MACHOs in this regime naturally results in repeating FRBs with
typical time delays of a few ms for a 
30\,$M_{\odot}$ lensing object. Searches for such events using existing instruments
(in particular \edit{CHIME/FRB}) could result in constraints on the fraction of dark matter
in MACHOs to better than 10\%.} 

\subsection{FRBs as probes of fundamental physics}\label{sec:4_6}

For most studies, the time delay between the radio signal from an FRB at different frequencies is interpreted as being \edit{primarily} due to plasma dispersion from ionized material along the line of sight. As discussed by \citet{wei_2015_phrvl}, from a fundamental physics perspective, additional contributions are possible due to Lorentz invariance violations, a non-zero rest mass of photons, and
violations of the Einstein Equivalence Principle. A violation of the latter would mean that photons of different energies fall at different rates when traversing gravitational potentials, rather like lunar laser ranging experiments probe the Earth and Moon falling in the Sun's gravitational potential \citep{nordvedt_1970_icar}. A non-zero photon mass results in a dispersion relationship in pulse propagation as a function of frequency. Failure to detect such a delay, allows a direct constraint on the photon mass.

Initial attempts to apply these ideas 
\citep{tingay_2016_apjl,wu_2016_apjl}, turned out to be thwarted due to being based upon FRB~20150418A which was not convincingly established to be associated with a galaxy at $z=0.492$ \citep{keane_2016_natur}. While subsequent constraints based on FRBs~20180924B and 20190523A, which have robustly determined host galaxies and redshift measurements, were found by \citet{wang_2020_pdu}, the current best \edit{limits} are found by \citet{xing_2019_apjl}, who consider the time delay between downward drifting (sad trombone) pulses in \aofrb. Using these pulses these authors limit equivalence principle violations to almost a part in $10^{16}$, while also placing a bound on the photon mass of $<5 \times 10^{-51}$\,kg. 

\section{Motivations for continued and future observations}\label{sec:futureObs}

Here we outline the various ways in which future observations can help decipher the FRB mystery (while, inevitably, leading to new questions as well).

\subsection{Much larger burst sample}\label{sec:5_1}

The coming 5 to 10 years of observations promise to deliver a sample of $>10,000$ FRBs, which will easily surpass the known Galactic population of radio pulsars found in the past half century.  Assuming that FRBs are related to neutron stars, we may thus soon know of more extragalactic neutron stars than those in our own Milky Way!  This large sample is critical for population synthesis studies \citep{james_2022_mnras_2}, and for applications like galaxy cross-correlation to measure the statistical properties of the local environments \citep{rafieiravandi_2021_apj} and IGM \citep{macquart_2020_natur}.  In fact, detailed studies of individual FRBs may be strongly limited by available time on other facilities, so statistical studies using archival optical/X-ray/gamma-ray survey data will be just as important as deep targeted studies of the most interesting individual sources.

With \edit{the availability of} more known FRBs, sub-dominant populations --- beyond the current repeater versus one-off dichotomy, and the 4 burst archetypes identified in the first CHIME/FRB catalogue \citep{pleunis_2021_apj} --- may slowly appear.  GRBs again provide an interesting precedent: beyond short and long gamma-ray bursts, it appears that some small fraction of \edit{the} observed GRBs are from magnetar flares like those seen from SGRs \citep{svinkin_2021_natur}.  In addition to the temporal, spectral and polarimetric properties of the FRBs \edit{themselves} \citep{pleunis_2021_apj}, sub-populations may be identifiable based on (periodic) activity level \citep{chime_2020_natur_582}, burst energy distribution \citep{shannon_2018_natur}, local environment \citep{tendulkar_2021_apjl} and coincidence with other multi-wavelength (or multi-messenger!) events \citep{mereghetti_2020_apjl}.

As more FRBs are found with timescales of nanoseconds to hundreds of milliseconds, and distances of kiloparsecs to gigaparsecs, we will gain a much clearer view of how the transient parameter space is populated by short-duration coherent transients \citep[see Fig.~\ref{fig:tps_zoom}; also][]{nimmo_2021_arxiv_arxiv210511446}.  Beyond aiding in identifying different sub-populations of source type or emission mechanism, this will inform where the community might best invest in future instrumentation and observing campaigns --- either to maximize detection yield or, conversely, to explore for new phenomena.  Current facilities are primarily focused on searching for FRBs at radio frequencies \edit{of} 0.4--1.7\,GHz and on timescales of $\sim100$\,$\upmu$s--$100$\,ms.  This may be giving us a highly biased view of the population.

\subsection{Long-term monitoring}\label{sec:5_2}

It will soon be a decade since the first-known burst from the first-known repeater, \aofrb, was detected by Arecibo \citep{spitler_2014_apj}.  Several CHIME/FRB, ASKAP, \edit{FAST} and Parkes\edit{-discovered} repeaters now also have multi-year spans of burst detections.  DM, RM and $\tau_{\rm scatt}$ timeseries will trace the dynamic local environments of some repeaters \citep{hilmarsson_2021_apjl}.  This can both inform progenitor models as well as whether the observed burst properties are predominantly intrinsic, or mainly reflect local propagation effects, \edit{like plasma lensing} \citep{cordes_2017_apj}.  We also need to determine whether the \edit{`sad trombone'} time-frequency drifting effect seen in repeaters \citep{hessels_2019_apjl} evolves with time: e.g., a systematic change in drift rate \edit{during activity periods}.  Identifying correlations in the burst properties of repeaters can inform the emission mechanism \citep{rajabi_2020_mnras}, and this needs to be done for a larger sample of sources.  Long-term monitoring of repeaters also opens interesting avenues for constraining cosmological parameters via gravitational lensing \citep{wucknitz_2021_aa}.

High-activity periods (`burst storms') will provide large burst numbers for statistical studies like burst energy distribution and wait times, and can also constrain models.  The baseline of apparently one-off FRBs detected by CHIME/FRB \citep{chime_2021_arxiv_arxiv210604352} is also important, since some of these sources may suddenly become very active compared to previous monitoring.  FRB~20201124A is currently the best example of this as it was seen to repeat actively after years of non-detection in CHIME/FRB monitoring of the sky \edit{\citep{Lanman_2021_arXiv}}.  On the other hand, continued monitoring will place increasingly stringent constraints on the repetition rate of apparent one-off FRBs.  For example, some apparently one-off FRBs may be associated with rare neutron star activity like the magnetar giant flares, which could potentially only recur once in decades for any individual source.

\subsection{Higher \edit{(and lower!)} time resolutions and wider bandwidths}\label{sec:5_3}

Observed FRB timescales extend over \edit{8} orders of magnitude from $\sim60$\,ns to $\sim3$\,s.  Nonetheless, there are still orders of magnitude to explore in the time dimension.  \edit{Future experiments and surveys should endeavor to push the envelope in order to search for both `ultra-fast radio bursts' (uFRBs; $\Delta t < 10$\,$\mu$s) and `not-so-fast radio bursts' (nsFRBs; $\Delta t > 100$\,ms).}

Nyquist-limited data can provide $\sim1$\,ns time resolution and can establish stronger observational links to phenomena seen from the \edit{young} Crab pulsar \citep{hankins_2016_apj} \edit{and `Crab twin' PSR~J0540$-$6919 \citep{geyer_2021_mnras} as well as the much older millisecond pulsars, like PSR~B1937+21 and PSR~B1821$-$24A, that emit giant pulses \citep{johnston_2004_iaus}.  These short timescales may also reveal new sub-populations of FRB sources and provide new avenues for using FRBs as astrophysical and cosmological probes.}  To avoid scattering from the Galactic foreground, this must be done at relatively high radio frequencies ($>2$\,GHz) and/or at high Galactic latitude \citep{nimmo_2021_arxiv_arxiv210511446}.  At comparably long timescales of $>100$\,ms, fast imaging searches will provide enhanced sensitivity compared to beam-formed techniques, since they provide a more stringent spatial filter.

While wide-band single-pixel feeds, and multi-band/multi-telescope campaigns, have better characterised the instantaneous spectra of FRBs \citep{kumar_2021_mnras}, there is still much to be explored: e.g., whether the \edit{broadband nature} of FRBs correlates with their energy or other burst properties \citep{hewitt_2021_arxiv,metzger_2021_arxiv}.  Such observations are also critical to determine whether the observed spectra are dominated by intrinsic or extrinsic effects.  Multi-telescope simultaneous observations from tens of MHz to several GHz (or more) can map the burst width and spectral variations to determine whether these are consistent with scattering and scintillation, or demonstrably unrelated to propagation effects.  

To date, there have been few attempts to detect FRBs at radio frequencies $>3$\,GHz \citep{michilli_2018_natur,gajjar_2018_apj}, but such efforts should increase since they provide access to the shortest-observable timescales and because the Crab pulsar demonstrates that a single source can produce remarkably different burst properties in different radio frequency bands \citep{hankins_2016_apj}.  Identifying FRBs at high frequencies can be challenging because the short dispersive delay makes it harder to distinguish from radio frequency interference and higher frequency observations are restricted to small fields of view.  Such observations can best be done using multiple geographically separated telescopes to verify the astrophysical nature of any detected events.

\subsection{A much larger host galaxy sample}\label{sec:5_4}

With close to \edit{two dozen} FRB hosts known, it has become apparent that FRBs can be found in a wide range of galaxy types with different total stellar mass, metallicity and star-formation rate \citep{heintz_2020_apj,mannings_2021_apj,bhandari_2021_arxiv}.  A much larger sample is needed, however, to look for preferences to certain galaxy types, which will also inform the possibility of FRB sub-populations with distinct physical origins.  Thus far, no FRB has been found coincident with a starburst galaxy, but perhaps some exceptionally young, energetic and active sources can be found in such environments.  Searches of nearby starburst galaxies, like M82, should continue since young and nearby FRBs may provide important multi-wavelength detections or constraints.  \edit{Here too, higher radio frequencies may be needed to peer into the depths of star forming regions where young FRBs may live.}

With a large ($>$100--1000) sample of host galaxies, the statistical properties of the local environments and host DM contribution will also be better known.  This is important for understanding how `cleanly' FRBs can serve as cosmological probes \citep{macquart_2020_natur}.  It may be the case that some FRBs are good probes of the IGM and CGM, while others are better probes of their local environment or reside in dense massive halos \citep{rafieiravandi_2021_apj,niu_2021_arxiv}.

\subsection{Zooming-in on the local environments of nearby FRBs}\label{sec:5_5}

In terms of characterising the local environments of FRBs, milliarcsecond localisations via VLBI only have added value, compared to $\sim$arcsecond localisations, for relatively nearby ($z\lesssim0.2$) sources where the host galaxy can be well resolved through photometry and spectroscopy by 8-m class ground-based facilities, or by {\it Hubble}.  Exceptionally nearby FRBs (within 100\,Mpc) are more rare \citep{bhardwaj_2021_apjl_910,bhardwaj_2021_apjl_919} but increasing the known sample to a few dozen will provide a major step in comparing local environments.  What fraction of FRBs are found in globular clusters \citep{kirsten_2021_arxiv}, what is their distribution with respect to local star forming regions \citep{tendulkar_2021_apjl}, and how many have compact persistent radio (or multi-wavelength) counterparts \citep{chatterjee_2017_natur,marcote_2017_apjl,niu_2021_arxiv,law_2021_arxiv}?  Within the next 5 years, the 39-m European Extremely Large Telescope (E-ELT) will provide angular resolution that surpasses that of {\it Hubble} by an order of magnitude.  Coupled with its enormous collecting area, this may allow us to associate some FRBs with stellar companions in a binary system, a plausible scenario given the periodic activity of some repeaters \citep{chime_2020_natur_582}.  The imminent launch of the long anticipated {\it James Webb Space Telescope} will provide another avenue for deep characterisation of FRB local environments.

\subsection{Targeted observations of potential FRB sites}\label{sec:5_6}

Linking FRBs to other known types of astrophysical sources and events can greatly inform their nature (and vice versa).  Targeted observations of recent (superluminous) supernova and long GRB sites may eventually lead to the discovery of ultra-young FRBs \edit{observed just after birth, if formed in such scenarios} \citep{eftekhari_2019_apjl,marcote_2019_apjl}.  \edit{We should also be targeting the population of `wandering' black holes in dwarf galaxies \citep{reines_2020_apj}, since some of these compact persistent radio sources may actually be related to repeating FRBs as opposed to AGN-like activity \citep{eftekhari_2020_apj,law_2021_arxiv}.}

Automated, low-latency follow-up as soon as possible after \edit{a transient} event \citep{anderson_2021_pasa} may reveal bursts from a central engine that later collapses to a black hole.  Co-observing, or buffered data from low-frequency radio telescopes (where the large dispersive delay offers the opportunity to capture the signal later), will detect or constrain prompt coherent radio emission from such events.  As the fields of view, and rapid, triggered repointing capabilities of radio telescopes increase, we may also find prompt FRB-like emission from cataclysmic events such as supernovae (SNe), tidal disruption events (TDEs), short/long GRBs, and gravitational wave (GW) events.

Targeted observations of magnetars, ULXs, microquasars, gamma-ray binaries, etc., may also yield important connections \citep{sridhar_2021_apj}.  Triggered observations are needed to catch these sources during active states; high-cadence monitoring is necessarily to find rare (but bright) events like the MJy-ms event from SGR~1935+2154 \citep{chime_2020_natur_587,bochenek_2020_natur}.  Because bright bursts are typically rare, and commonly occurring bursts are typically faint, a tiered approach, using multiple instruments with different trade-offs between sensitivity and on-sky time \edit{is} needed.  CHIME/FRB provides daily monitoring of the northern hemisphere, but with only $\sim10$\,min integration in the main beams.  This can be complemented by low-sensitivity arrays, like STARE2 and the proposed GRex, that are observing (close to) the whole sky simultaneously \citep{connor_2021_pasp}, as well as high-cadence observations of known sources using 25-m class radio telescopes \citep{kirsten_2021_arxiv}.  The Galactic centre and bulge deserve particular attention; high-cadence X-ray monitoring has demonstrated the scientific potential of this approach \citep{degenaar_2010_aa}.

\subsection{Multi-wavelength and multi-messenger connections}\label{sec:5_7}

Thus far, extragalactic FRBs have appeared quiet at other wavelengths \citep{scholz_2017_apj,scholz_2020_apj} and via other messengers \citep{aartsen_2020_apj}.  It is possible that the distances to FRBs are simply too large \edit{for there to be detectable multi-wavelength counterparts, given} the sensitivities of existing telescopes observing above the radio band.  We have only scratched the surface, however, since most multi-telescope studies have targeted repeaters and prompt observation of one-off FRBs is particularly challenging.  Future FRB search machines would benefit from dedicated optical and high-energy instruments simultaneously observing the same field of view.  

Assuming a scenario where FRBs originate from neutron star magnetospheres, \citet{lyutikov_2016_apjl} consider what electromagnetic counterparts might accompany them.  In the case of flares associated with magnetospheric reconfigurations, contemporaneous bursts at other wavelengths, and/or an afterglow may be detectable.  They argue that a short GRB-like burst, a millisecond-duration optical flash, and/or a high-energy afterglow are the most promising scenarios to pursue observationally.  Radio interferometers like ASKAP that can both discover and localise one-off FRBs, can also place limits on precursor and afterglow radio emission \citep{bhandari_2020_apjl_901}.  

\subsection{Linking to other observed radio phenomena}\label{sec:5_8}

We may also come to a deeper understanding of FRBs by studying potentially related source types and emission processes.  Though they have been studied for over a half century, there is still no broad consensus on the mechanism(s) by which radio pulsars produce their emission.  Nonetheless, advanced numerical simulations are providing new insights into pulsar emission and how neutron star magnetospheres can create electromagnetic phenomena in general \citep{philippov_2020_phrvl}.  Similar simulations focusing specifically on reproducing the phenomenology and much higher luminosities of FRBs are needed.

At the same time, we continue to discover new emission phenomena from Galactic pulsars and magnetars \citep{bij_2021_apj,lower_2021_mnras}. 
High-cadence observations of pulsar single pulses may reveal new phenomena that also \edit{give} insight into FRBs.  Wide-field FRB machines like CHIME/FRB, and high-sensitivity, high-cadence pulsar observations with MeerKAT can elucidate this.  High-time-resolution studies of giant pulses from young pulsars and millisecond pulsars deserve increased attention in the context of understanding FRB emission.  Lastly, continued studies of short-timescale radio emission from the Sun, active/flare stars, and (exo)planets are likely to reveal new phenomena and may provide interesting analogies for understanding FRBs.

\section{Future prospects}\label{sec:FutureProspects}

\edit{In this update to our 2019 review, we have focused on recent observational highlights in the rapidly evolving study of FRBs. These observations shed light on new aspects of the FRB phenomenon, constrain theoretical models from new angles, and show early promise for using FRBs as astrophysical and cosmological probes. As with our previous review, we anticipate that many elements of the summary presented here will fall rapidly out of date.  This underscores the rapid rate of progress that is being made in the field!} 

\edit{Many aspects of the FRB phenomenon remain to be explored. Clear areas for progress identified while writing this review include: expanded searches for FRBs at high ($>$2 GHz) and low ($<$300 MHz) radio frequencies; higher time and frequency resolution exploration of burst and sub-burst morphology; and searches for periodic activity in a larger sample of regularly repeating sources. The future also holds promise at other wavelengths, including: the search for X-ray and optical counterparts associated with FRBs; deep optical imaging of host galaxies and local environments; rapid triggering of follow-up telescopes through FRB VOEvents; and coordinated campaigns with other facilities for targeting periodically active repeaters.} 

\edit{Existing radio facilities ramping up their search efforts provide complementary views of the FRB phenomenon: e.g., from the shallow, wide-field efforts of STARE2 \citep{bochenek_2020_natur} to the pinpoint precision of the EVN \citep{marcote_2020_natur} and the CHIME/FRB Outriggers \citep{leung_2021_aj}. A large sample of FRBs from telescopes around the world searching at different radio frequencies, sensitivities, time resolutions, and fields of view will offer more insight into the menagerie of FRB properties. Future facilities such as CHORD, DSA-2000, and the SKA are expected to build on these successes in the coming decade.} 

\edit{The future is indeed (radio) bright. As with our previous review, we conclude with our predictions for the state of the field five years from now.}
    
\subsection{Predictions for \edit{2027}}\label{sec:predictions}
\subsubsection{EP}

Looking back at my predictions from our previous review, it is easy to remember the state of the field at that time; the discovery space felt wide open for surprises. Some of my predictions have already come true -- FRBs over many decades of radio frequency, FRBs in nearby ($\sim$Mpc) galaxies, a large population from wide-field arrays. Some predictions have yet to be realised -- a high-redshift host, a better understanding of FRB polarisation -- and some of the most exciting results from the past few years surprised us all, like the dazzlingly bright burst from SGR~1935+2154. In the next $\sim$5 years I predict we will see more FRB-like radio behavior from Galactic magnetars and establish a \edit{firm} FRB---magnetar connection for at least a subset of the FRB population. I predict we will have over 100 host galaxies through the efforts of ASKAP, and other precision interferometers \edit{with the sample increasingly dominated by the tens of host galaxies per year from CHIME/FRB Outriggers}. I predict that we will see more FRBs at low ($\sim$100\,MHz) radio frequencies and we will also see FRBs at high ($\geq10$\,GHz) frequencies, if we look for them. I predict that we will identify two distinct populations -- FRBs that repeat and FRBs that do not -- and morphology will be an important predictor of an FRB source's probability of repeating. I am hopeful that a small sample of sources in nearby galaxies will give us insight into the environment, activity level, and multi-wavelength emission of FRBs that can be extended to our understanding of sources at greater distances. As before, I predict that our vibrant and diverse community will continue to grow and benefit from \edit{the} expertise of observers and theorists from a wide range of astrophysical backgrounds. I look forward to seeing how right or wrong these predictions are found to be, and I also look forward to the next big surprises these curious sources have to offer us!

\subsubsection{JWTH}

Re-reading my predictions from the original version of our review in 2019, I can see that many of these things have come to pass, already in the last 2 years!  This speaks volumes about the rate at which the field is progressing.  I should have been more ambitious and wilder in my predictions.  But there have also been many discoveries that we did not foresee at the time of our original review --- like periodic activity in repeaters and a globular cluster host for one nearby repeater.  Most likely the safest prediction to make, moving forward, is that there are more surprises to come.  As hoped, we have now detected (sub-)microsecond structure in FRBs.  This has only been possible, so far, in follow-up observations of repeaters.  I predict that in the next 5 years we will start blindly detecting `ultra-fast radio bursts' (uFRBs) in searches that employ high radio frequencies and semi-coherent dedispersion trials.  In a more general sense, I think that the transient phase space will continue to be filled-in on the shortest-observable timescales, and for the rarest forms of astrophysical events.  \edit{Conversely, I also think that there are great prospects for uncovering a population of `not-so-fast radio bursts' (nsFRBs) with durations of 100\,ms to several seconds.}  Clusters in \edit{the short-duration transient} phase space might hint at distinct populations, but given that some neutron stars (e.g., SGR~1935+2154) can produce radio bursts spanning many orders of magnitude in timescale and luminosity, this will require additional contextual information to justify any such groupings.  Going more out on a limb, I predict that we {\it will not} detect prompt X-ray or gamma-ray counterparts to {\it extragalactic\footnote{The Galactic magnetar SGR~1935+2154 suggests that we should see this for extremely nearby sources, however.}} FRBs (or vice-versa) within the next 5 years --- though I do hold more hope for the detection of prompt optical flashes.  By targeting the sites of known astrophysical explosions (various flavours of supernovae and kilonovae), I predict that we will directly link a longer-lived, repeating FRB source to the event that created the central engine.

\subsubsection{DRL}

As we finalise this review, there are a number of exciting results emerging and it is clear that the CHIME/FRB sample and other experiments are going to guarantee a number of surprises in the coming five years. I am gratified my 2019 prediction, from our original review, that only 1\% of all FRBs will show repeats is not too far off! At the time of writing, about 4\% of the CHIME/FRB sample repeats. I predict this fraction will go down further as the survey runs out of repeating sources to find down to its flux limit. My prediction of 3000 FRBs by 2024 looks like it will easily be reached at the current pace of discoveries. With the plans for CHIME/FRB outrigger stations now well underway, and progress at many other sites, there are now very good reasons to be optimistic about getting a large number (perhaps 1000) robustly
associated galaxies with FRBs over the next five years. For me, one of the most interesting of the current results are the morphological studies now being provided by the first CHIME/FRB catalogue. The clearly distinct classes emerging from this sample indicate to me strong evidence in favor of multiple source populations. We are clearly just scraping the surface right now, but I will boldly predict that a single source population for FRBs will no longer be viable by the middle of this decade. \edit{I am optimistic that prompt electromagnetic counterparts will soon be observed for some FRBs in the form of optical flashes. Spurred on by Bing Zhang, I also dare to dream of an association between an FRB and a gravitational wave source.}

\begin{acknowledgements}
The authors would like to dedicate this review to the late J.-P. Macquart, an esteemed friend, colleague, and mentor. Thank you, J.-P., for your insights and boundless enthusiasm for all things FRB-related, and beyond.

We thank Bridget Anderson, Anya Bilous, Joeri van Leeuwen, Kenzie Nimmo, Leon Oostrum, In\'{e}s Pastor-Marazuela, and Kendrick Smith for their feedback on drafts of this review. \edit{Bing Zhang, Jonathan Katz, Julian Mu\~noz, Daniele Michilli, Sandro Mereghetti, Shami Chatterjee and Mark Snelders provided much useful feedback on our initially submitted version of the manuscript.}
\edit{We additionally thank the two anonymous referees who improved the quality and clarity of the manuscript.}

EP is supported by a \edit{Dutch Research Council} (NWO) Veni Fellowship.
JWTH is supported by an NWO Vici grant (`AstroFlash'; VI.C.192.045).
DRL acknowledges support from the Research Corporation for Scientific Advancement and
the National Science Foundation through award AAG-1616042.

\end{acknowledgements}

\bibliographystyle{spbasic-FS}
\bibliography{frbrev2}

\end{document}